%% file: arxiv.tex
\documentclass{article}

\usepackage[preprint]{neurips_2025_libribrain_competition}

\usepackage[utf8]{inputenc} 
\usepackage[T1]{fontenc}    
\usepackage[hidelinks]{hyperref}       
\usepackage{graphicx}
\usepackage{amsmath}
\usepackage{amsfonts}
\usepackage{fancyhdr}
\usepackage{url}            
\usepackage{caption}
\usepackage{subcaption}
\usepackage{booktabs}
\usepackage{float}
\usepackage{pifont}         

\newcommand{\xmark}{\ding{55}} 

\title{Measuring Robustness of Speech Recognition from MEG Signals Under Distribution Shift}
\author{
  Sheng-You Chien \\
  Department of Computer Science \\
  National Tsing-Hua University \\
  Hsinchu, Taiwan \\
  \texttt{s99086tobby@gmail.com}
  \AND
  Bo-Yi Mao \\
  Department of Computer Science \\
  National Tsing-Hua University \\
  Hsinchu, Taiwan \\
  \texttt{dogeon188@gapp.nthu.edu.tw}
  \AND
  Yi-Ning Chang \\
  Department of Computer Science \\
  National Tsing-Hua University \\
  Hsinchu, Taiwan \\
  \texttt{changyn@gapp.nthu.edu.tw}
  \AND
  Po-Chih Kuo \\
  Department of Computer Science \\
  National Tsing-Hua University \\
  Hsinchu, Taiwan \\
  \texttt{kuopc@cs.nthu.edu.tw}
}
\date{}  

\begin{document}

\maketitle


\input{nstc/abstract-eng}

\input{nstc/introduction}

\input{nstc/related_works}

\input{nstc/methodology}

\input{nstc/results}

\input{nstc/conclusion}

\input{nstc/availability}

\newpage


{\small \bibliography{milestone}}

\newpage

\appendix

\input{nstc/full_results}

\end{document}

%% file: nstc/abstract-eng.tex
\begin{abstract}
This study investigates robust speech-related decoding from non-invasive MEG signals using the LibriBrain phoneme-classification benchmark from the 2025 PNPL competition. We compare residual convolutional neural networks (CNNs), an STFT-based CNN, and a CNN--Transformer hybrid, while also examining the effects of group averaging, label balancing, repeated grouping, normalization strategies, and data augmentation. Across our in-house implementations, preprocessing and data-configuration choices matter more than additional architectural complexity, among which instance normalization emerges as the most influential modification for generalization. The strongest of our own models, a CNN with group averaging, label balancing, repeated grouping, and instance normalization, achieves 60.95\% F1-macro on the test split, compared with 39.53\% for the plain CNN baseline. However, most of our models, without instance normalization, show substantial validation-to-test degradation, indicating that distribution shift induced by different normalization statistics is a major obstacle to generalization in our experiments. By contrast, MEGConformer maintains 64.09\% F1-macro on both validation and test, and saliency-map analysis is qualitatively consistent with this contrast: weaker models exhibit more concentrated or repetitive phoneme-sensitive patterns across splits, whereas MEGConformer appears more distributed. Overall, the results suggest that improving the reliability of non-invasive phoneme decoding will likely require better handling of normalization-related distribution shift while also addressing the challenge of single-trial decoding.
\end{abstract}

%% file: nstc/introduction.tex
\section{Introduction}
\label{sec:introduction}

\subsection{Background and Motivation}
\label{subsec:background}

Decoding speech-related information from non-invasive brain recordings is an important problem for both cognitive neuroscience and brain--computer interfaces. Scientifically, it can help reveal how the brain processes speech and language; practically, it may support assistive communication technologies for people with severe motor or speech impairments. At the same time, the problem remains difficult because non-invasive neural signals are noisy, high-dimensional, and variable across subjects, sessions, and preprocessing pipelines. As a result, progress depends not only on model capacity but also on whether decoding methods remain robust under realistic distribution shifts.

The LibriBrain benchmark and the 2025 PNPL competition provide a useful setting for studying this challenge \citep{ozdogan2025libribrain,landau2025pnpl}. In LibriBrain, each example is a short MEG segment centered on a phoneme, which makes it possible to evaluate decoding methods under a standardized protocol. Compared with direct text or semantic reconstruction, phoneme classification is a narrower and more controlled task, but it still captures a meaningful level of speech-related information. This benchmark therefore offers a practical intermediate target for analyzing how preprocessing strategies, data augmentation, and model architecture affect decoding performance.

Motivated by this setting, our work focuses on phoneme classification on LibriBrain and examines how different design choices behave when the evaluation distribution differs from the development data. We compare residual convolutional models, STFT-based models, and CNN--Transformer hybrids, while also analyzing the roles of group averaging, feature standardization, and data augmentation. 

\subsection{Objectives}
\label{subsec:objectives}

The objectives of this project are as follows:

\begin{enumerate}
  \item To build a phoneme-classification pipeline for LibriBrain MEG data, including splitting, preprocessing, augmentation, and evaluation.
  \item To compare multiple neural architectures and determine how preprocessing and modeling choices influence decoding performance.
  \item To analyze the impact of distribution shift between validation and hidden evaluation data, especially the effect of different standardization strategies.
  \item To identify which modeling and preprocessing choices improve robustness to distribution shift, and which ones mainly enhance validation performance without generalizing to hidden evaluation conditions.
  \item To summarize the practical challenges encountered in this study and discuss how the findings can support future research on robust non-invasive speech decoding from MEG.
\end{enumerate}

%% file: nstc/related_works.tex
\section{Related Work}
\label{sec:related_works}

\subsection{Neural Decoding from Non-Invasive Brain Signals}
\label{subsec:noninvasive_decoding}

Compared with invasive neural interfaces, non-invasive brain recording modalities such as EEG and MEG are safer and easier to deploy, but their lower signal-to-noise ratios and weaker spatial specificity make robust decoding more challenging \citep{lotte2018,baillet2017,hamalainen1993magnetoencephalography,schirrmeister2017deep}.

Among these modalities, magnetoencephalography (MEG) is particularly attractive for speech-related decoding because it provides millisecond-level temporal resolution and directly captures fast neural dynamics associated with speech perception and processing \citep{hamalainen1993magnetoencephalography,baillet2017,lotte2018}. Prior work has used MEG to study auditory attention and speech perception from neural activity \citep{o2015attentional,defossez2023decoding}. Recent advances in deep learning, especially convolutional neural networks, have further improved the modeling of complex neural time series, making MEG-based speech decoding a useful testbed for robust neural decoding under realistic data and preprocessing constraints \citep{schirrmeister2017deep,defossez2023decoding}.

\subsection{Deep Learning for Neural Time-Series Modeling}
\label{subsec:timeseries_modeling}

Deep learning is widely used for neural time-series analysis because it can learn hierarchical temporal features from raw or lightly processed signals, and in multichannel recordings it can also exploit spatiotemporal structure \citep{gamboa2017deep,ismail2019deep,walther2023systematic}. In the broader time-series literature, recurrent models such as LSTMs, convolutional models, and hybrid recurrent--convolutional architectures have been used to model both long-range dependencies and local patterns \citep{karim2017lstm,zhao2017convolutional,elsayed2018deep}. Transformer-based EEG models extend this line of work by using self-attention to capture longer-range context \citep{sun2021eeg,vafaei2025transformers}. These modeling choices are relevant to MEG decoding, where informative patterns may be distributed across channels and unfold over short time windows.

\subsection{Speech and Phoneme Decoding from MEG}
\label{subsec:speech_phoneme_decoding}

Early non-invasive speech-decoding work showed that MEG can discriminate phonemic contrasts and estimate stimulus timing from brief neural responses \citep{lukka2000}. Later MEG and EEG studies extended decoding to spoken and imagined phrases and to phoneme-level classification, indicating that speech-related information is present in both modalities \citep{dash2020decoding,larocco2023evaluation}. Recent benchmarks such as LibriBrain provide a standardized MEG setting for phoneme classification at scale \citep{ozdogan2025libribrain,landau2025pnpl}. Related work also uses phonological or articulatory features as intermediate targets, offering a structured alternative to direct phoneme labels \citep{wang2012using,moreira2025open}. This motivates our focus on phoneme classification rather than direct text generation.

This direction choice is also supported by recent critiques of open-vocabulary EEG-to-text decoding. A recent reanalysis of transformer-based EEG-to-text systems on the ZuCo datasets showed that reported performance can be substantially inflated by teacher-forcing during evaluation \citep{jo2024eeg}. The same study further found that several seq2seq models performed similarly on random noise and EEG, suggesting that end-to-end text reconstruction may reflect memorization of linguistic priors or labels rather than reliable extraction of neural speech content. This strengthens the case for using constrained targets such as phonemes or phonological features when evaluating whether non-invasive brain signals contain decodable speech information.

\subsection{Domain Adaptation and Distribution Shift}
\label{subsec:domain_shift}

Distribution shift is a major challenge in brain signal decoding due to variability across subjects, recording sessions, and preprocessing pipelines. Models trained on one split, subject, or pipeline may fail to generalize to another. Related work in adjacent domains has explored teacher--student distillation for speech recognition, MMD-based domain alignment, and adversarial or self-supervised approaches in EEG-based settings to reduce this mismatch \citep{meng2019teacherstudent,gretton2012kernel,lin2023multi,wang2025mmoc}.

Recent work on speech representations in non-invasive brain recordings likewise identifies domain shift as a central obstacle to robust decoding across recording conditions \citep{ridge2024,jayalath2025}. For this reason, we treat normalization and data augmentation as part of the adaptation problem rather than as purely technical details; in related speech settings, augmentation has been shown to improve robustness to nuisance variability and train--test mismatch \citep{park2019specaugment}.

%% file: nstc/methodology.tex
\section{Methodology}
\label{sec:methodology}

\subsection{Dataset and Preprocessing}
\label{subsec:dataset}

\subsubsection{Dataset Overview}

The LibriBrain dataset \citep{ozdogan2025libribrain} is a large-scale, naturalistic MEG dataset collected from a single human subject while they listened to the Sherlock Holmes audiobook series. The dataset includes 93 recording sessions, each correspond to a chapter of the audiobook. The dataset is partitioned into training, validation, test, and holdout splits, with the holdout split containing data from the subject not included in the public dataset and used for final evaluation in the PNPL competition.

\begin{figure}[h]
  \centering
  \includegraphics[width=0.85\linewidth]{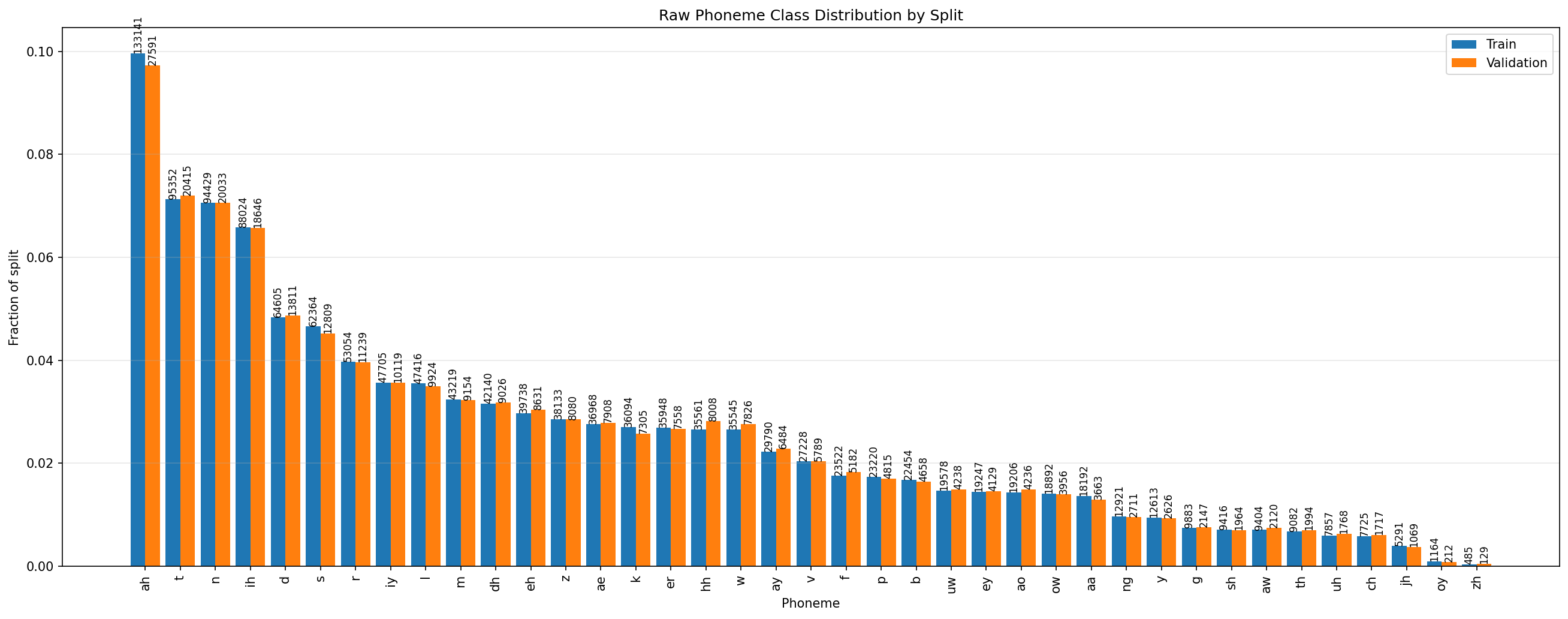}
  \caption{Raw phoneme-class distribution for the train and validation splits.}
  \label{fig:phoneme_class_distribution}
\end{figure}

Each sample in the dataset is a 500\,ms MEG segment centered on a phoneme, recorded from 306 channels at 1\,kHz, and then downsampled to 250\,Hz. The phoneme inventory consists of 39 classes, and the dataset exhibits a long-tailed distribution of phoneme frequencies, as shown in Figure~\ref{fig:phoneme_class_distribution}.

The dataset is available via the Python library \texttt{pnpl} and available for download from Hugging Face\footnote{\url{https://huggingface.co/datasets/pnpl/LibriBrain}}.

\subsubsection{Dataset Partitioning}

For training and evaluation, we use the LibriBrain phoneme classification benchmark from the PNPL competition, but with dataset splits that differ from those in the default PNPL library. The full dataset contains 91 trials and 1,622,678 phoneme instances. Table~\ref{tab:splits} summarizes the number of trials, phoneme instances, and standardization strategy for each split. Our validation set consists of 14 trials from subject 0: sessions 11 and 12 for Sherlock1--4 (trial 2 for Sherlock1 and trial 1 for Sherlock2--4), sessions 14 and 15 for Sherlock5 (trial 1), and sessions 13 and 14 for Sherlock6--7 (trial 1). Features are standardized separately for each split: the training and validation sets use statistics computed from the training set, whereas the test and holdout sets use their own split-specific statistics.

\begin{table}[t]
  \centering
  \caption{Summary of LibriBrain phoneme classification dataset splits.}
  \label{tab:splits}
  \begin{tabular}{lccc}
    \toprule
    Split & Trials & Phonemes & Standardization \\
    \midrule
    Train      & 76 & 1,336,606 & Train statistics \\
    Validation & 14 &   283,690 & Train statistics \\
    Test       & 14 &   283,690 & Test statistics \\
    Holdout    &  1 &     2,382 & Holdout statistics \\
    \midrule
    Total      & 91 & 1,622,678 & -- \\
    \bottomrule
  \end{tabular}
\end{table}

\subsubsection{Preprocessing}
\paragraph{Preprocessing Pipeline.}
The LibriBrain dataset is released with a minimal preprocessing pipeline applied to preserve signal quality while keeping the data accessible for downstream machine learning research \citep{ozdogan2025libribrain}. Head position information is first used for motion correction, and bad channels are identified for interpolation. External and stationary noise is then reduced with Maxwell filtering based on signal space separation (SSS). Additional notch filters remove 50\,Hz line noise and its 100\,Hz harmonic. The recordings are subsequently band-pass filtered between 0.1 and 125\,Hz using zero-phase two-pass Butterworth filters to reduce slow drifts and aliasing artifacts while retaining high-frequency information. Finally, the signals are downsampled from 1\,kHz to 250\,Hz and released in serialized HDF5 format for efficient use in deep learning pipelines.

\paragraph{Group averaging.}
We apply within-class window averaging to improve the signal-to-noise ratio. Unless otherwise specified, we randomly group 100 samples from the same phoneme class and replace them with their average. In the ablation study, we vary the number of repeated samplings per epoch and report results for repeat counts of 1, 5, and 10.

\paragraph{Label balancing.} To mitigate the long-tailed class distribution, we apply random oversampling to the training set, ensuring that each phoneme class has the same number of samples as the most frequent class. This is implemented by randomly duplicating samples from underrepresented classes until all classes have equal representation in the training data.

\subsubsection{Data Augmentation}
\label{subsubsec:augmentation}
To improve robustness, we apply stochastic data augmentation during training. Each augmentation is applied independently with probability 0.3. The augmentation strategies are as follows:

\begin{itemize}
  \item \textbf{Gaussian noise:} additive noise with standard deviation equal to 0.01 times the sample standard deviation.

  \item \textbf{Temporal shift:} circular time shifts uniformly sampled within $\pm 40$\,ms.

  \item \textbf{Temporal masking:} zeroing out up to 80\,ms of contiguous temporal samples.

  \item \textbf{Channel dropout:} randomly zeroing 10\% of MEG sensors per training step.

  \item \textbf{Amplitude scaling:} multiplying the waveform by a factor drawn from $U(0.9, 1.1)$.

  \item \textbf{Frequency band perturbation:} performing a Fourier transform on the input signal (up to 100\,Hz), randomly selecting one or more frequency bands, scaling their amplitudes by a factor sampled from $[0.8, 1.2]$, and transforming the modified spectrum back to the time domain.
\end{itemize}

\subsection{Model Architecture}
\label{subsec:model_architecture}

\subsubsection{Instance Normalization}

To mitigate distribution shift between the training and holdout splits, we apply instance normalization \citep{ulyanov16instnorm} to each sample independently, which normalizes each channel to zero mean and unit variance based on its own statistics. For a group-averaged sample $x \in \mathbb{R}^{C \times T}$ with $C$ channels and $T$ time points, the normalized sample $\hat{x}$ is computed as Equation~\ref{eq:instnorm}:

\begin{equation}
  \label{eq:instnorm}
  y = \frac{x - \mathbb E[x]}{\sqrt{\mathrm{Var}[x]}} \cdot \gamma + \beta,
\end{equation}

where $\mathbb E[x]$ and $\mathrm{Var}[x]$ are the mean and variance of $x$ computed per-window and per-channel over the time dimension; $\gamma = 1$ and $\beta = 0$ are fixed, following the implementation of \citep{dezuazo2025megconformer}. 

This approach is motivated by the observation that the holdout split exhibits a different global amplitude distribution from the training set, likely due to differences in recording conditions or subject state. By normalizing each sample separately, we aim to reduce this shift while preserving relative patterns across channels and time.

\subsubsection{Architectures}

\begin{figure}[t]
  \centering
  \includegraphics[width=\linewidth]{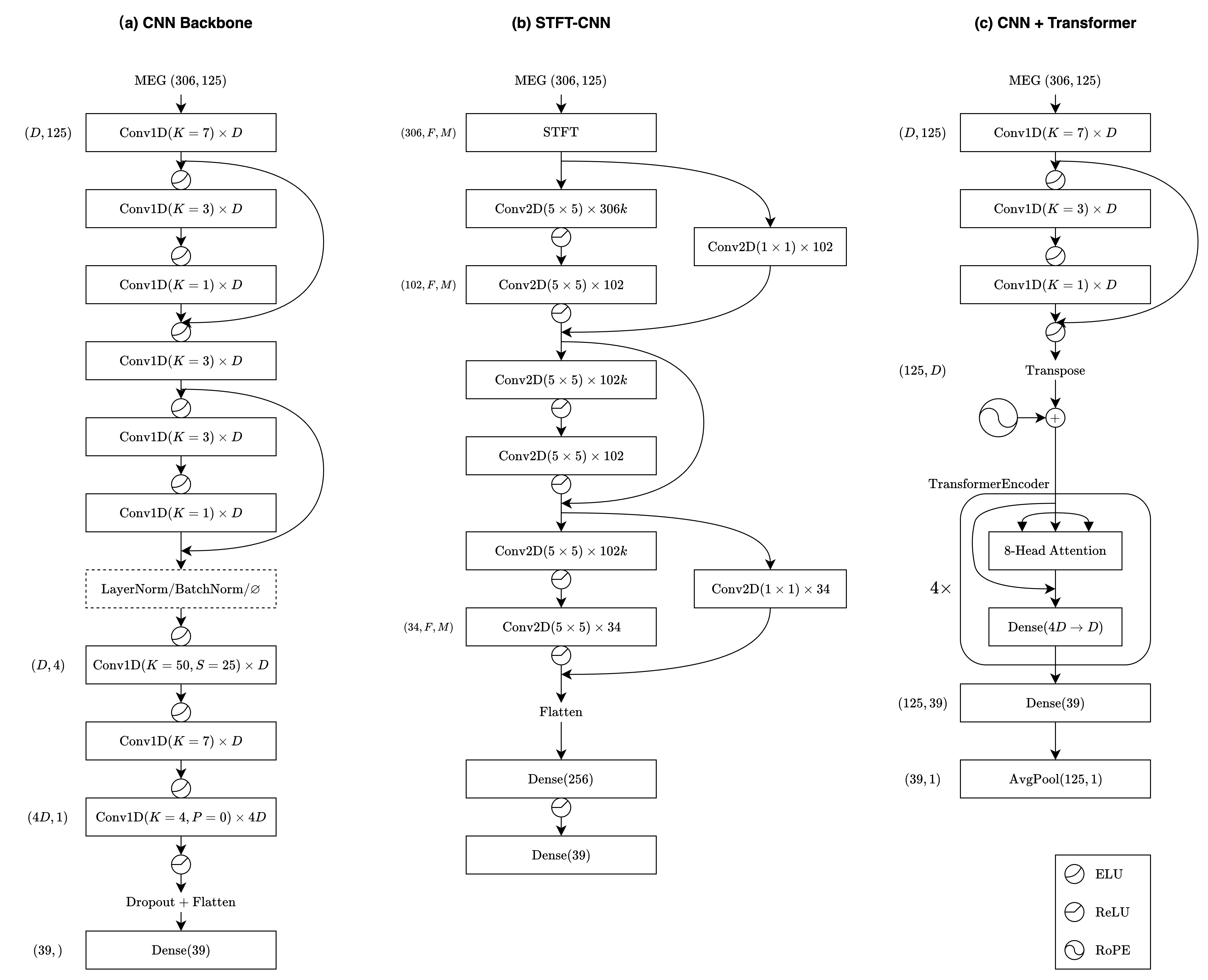}
  \caption{Overall model architecture. (a) CNN backbone. (b) STFT-CNN. (c) CNN-Transformer hybrid. For all convolutional blocks, the stride $S$ is set to $1$, and padding $P$ is chosen to preserve the temporal resolution unless otherwise specified.
  }
  \label{fig:architecture}
\end{figure}

We implement and evaluate three neural architectures, summarized in Figure~\ref{fig:architecture}.

\paragraph{ResNet-style CNN.}
Our primary architecture follows the baseline of Özdogan et al.~\citep{ozdogan2025libribrain}, which stacks temporal convolutional blocks with residual connections and group normalization. The backbone mainly consists of several 1D convolutional layers with channel dimension $D = 256$ and two residual connections. As shown in Figure~\ref{fig:architecture}(a), we optionally insert an additional normalization layer, either layer normalization or batch normalization. A lightweight classifier head then produces phoneme logits and, when enabled, phonological feature predictions.

\paragraph{STFT-based CNN.}
In this variant, we first apply a short-time Fourier transform (STFT) independently to each channel, using a window size of $25$ and a hop size of $5$. The resulting time--frequency representations are then processed by a shared 2D ResNet-style CNN, as illustrated in Figure~\ref{fig:architecture}(b). This design aims to exploit spectral structure while maintaining parameter sharing across channels.

\paragraph{CNN--Transformer Hybrid.}
To model longer-range dependencies across sensors and time, we extend the CNN backbone with a 4-layer Transformer encoder with 8 attention heads, as shown in Figure~\ref{fig:architecture}(c). The convolutional layers act as an initial filtering stage before the Transformer captures broader interactions in the latent representation.

\paragraph{MEGConformer.}

MEGConformer \citep{dezuazo2025megconformer} is a Conformer-based neural decoder for the LibriBrain competition that uses raw 306-channel MEG signals as input. The model starts with a lightweight 1D convolutional projection layer that maps the sensor signals to a 144-dimensional latent space, followed by dropout, a stack of Conformer blocks, and a task-specific linear classifier. For phoneme classification specifically, the model uses 0.5\,s windows, a smaller custom Conformer with 7 layers and 12 heads, instance-level normalization to reduce holdout distribution shift, inverse-square-root class weighting to address label imbalance, and dynamic grouping of 100-sample averages during training. We include MEGConformer as an external baseline as it achieved state-of-the-art performance on the LibriBrain phoneme classification competition.

\subsection{Training Procedure}
\label{subsec:training_procedure}

The training procedure for our in-house models is implemented in PyTorch and conducted on a single RTX 5090 GPU. Unless otherwise noted, all architectures use the same training hyperparameters: AdamW with a learning rate of $10^{-4}$, a weight decay of $10^{-2}$, a batch size of 256, 10 training epochs, and group averaging with batches of 100 windows.

Model selection is based on validation F1-macro, and the checkpoint with the best validation score is used for final evaluation on the test and holdout splits.

Architecture-specific changes are limited to the corresponding design choices, such as inserting layer normalization or batch normalization, replacing the time-domain encoder with the STFT or Transformer variants, or optimizing PanPhon targets with an MSE objective followed by nearest-neighbor projection back to the phoneme inventory at evaluation time.

\subsection{Evaluation Metrics}
\label{subsec:evaluation_metrics}

Consistent with the evaluation protocol of the LibriBrain 2025 PNPL competition, we use macro-averaged F1-score (\textbf{F1-macro}) as the primary evaluation metric. This metric is appropriate for phoneme classification because it weights all 39 phoneme classes equally, regardless of their frequency in the dataset. It is therefore well suited to the long-tailed class distribution of naturalistic speech and provides a stricter measure of whether performance generalizes across development and hidden evaluation splits.

\subsection{Saliency Map}
\label{subsec:saliency_map}

\subsubsection{Saliency Score Computation}

To analyze which parts of each network are most sensitive to different phoneme classes, we generate layer-wise saliency clustermaps from the best saved checkpoints, with the resulting validation and test visualizations shown later in Figure~\ref{fig:saliency_map_grid}.

For a given split, we disable grouped-sample averaging and run the model on individual phoneme-centered windows. We register forward hooks on all trainable submodules with observable tensor outputs and, for each minibatch, compute the gradient of the target logit with respect to the activation of each hooked submodule. Let $a_l^{(i)}$ denote the activation tensor of sublayer $l$ for sample $i$, and let $z_{y_i}^{(i)}$ be the logit of the ground-truth phoneme class $y_i$. The per-sample saliency score $s_l^{(i)}$ is computed as the mean absolute value of the gradient, as defined in Equation~\ref{eq:saliency_score}:

\begin{equation}
  \label{eq:saliency_score}
  s_l^{(i)} = \frac{1}{|a_l^{(i)}|}\sum_{u \in a_l^{(i)}} \left| \frac{\partial z_{y_i}^{(i)}}{\partial a_{l,u}^{(i)}} \right|,
\end{equation}

where $u$ indexes all elements of the activation tensor and $|a_l^{(i)}|$ is the number of those elements. We then average these per-sample scores over all samples belonging to the same phoneme class $c$, as defined in Equation~\ref{eq:saliency_avg}:

\begin{equation}
  \label{eq:saliency_avg}
  S_{l,c} = \frac{1}{|\mathcal{I}_c|}\sum_{i \in \mathcal{I}_c} s_l^{(i)},
\end{equation}

producing a raw matrix $S$ whose rows correspond to model sublayers and whose columns correspond to phonemes. For visualization, each row of this matrix is independently min-max normalized to the range $[0, 1]$ before plotting. This normalization enables comparison of phoneme selectivity within each layer, rather than absolute saliency magnitude across different layers. We apply hierarchical clustering to both rows and columns using Euclidean distance and average linkage, which groups together sublayers with similar phoneme-sensitivity profiles and phonemes that elicit similar saliency patterns across the network.

\subsubsection{Cross-Split Saliency Similarity}

To quantify the consistency of saliency patterns between the validation and test splits, we compute layer-by-phoneme similarity matrices using Pearson and Spearman correlations on paired validation and test samples (i.e., the same sample with different standardization). The resulting statistics are shown in Table~\ref{tab:saliency_similarity} and visualized in Figure~\ref{fig:saliency_similarity_grid}.

For each layer $l$ and paired sample $i$, we first compute the scalar saliency score $s_l^{(i)}$ defined in the previous subsection. For each phoneme class $c$, we then collect the paired validation and test saliency-score sequences $\{(s_l^{(i,\mathrm{val})}, s_l^{(i,\mathrm{test})})\}_{i \in \mathcal{I}_c}$ over all samples in that class.

We then compute the Pearson correlation coefficient $r_{l,c}$ and Spearman rank correlation coefficient $\rho_{l,c}$ between these two saliency-score sequences for each layer $l$ and phoneme class $c$, as defined in Equations~\ref{eq:pearson} and \ref{eq:spearman}:

\begin{equation}
  \label{eq:pearson}
  r_{l,c} = \mathrm{corr}_{\mathrm{Pearson}}\!\left(\{s_l^{(i,\mathrm{val})}\}_{i \in \mathcal{I}_c}, \{s_l^{(i,\mathrm{test})}\}_{i \in \mathcal{I}_c}\right),
\end{equation}

\begin{equation}
  \label{eq:spearman}
  \rho_{l,c} = \mathrm{corr}_{\mathrm{Spearman}}\!\left(\{s_l^{(i,\mathrm{val})}\}_{i \in \mathcal{I}_c}, \{s_l^{(i,\mathrm{test})}\}_{i \in \mathcal{I}_c}\right).
\end{equation}

This yields one Pearson matrix and one Spearman matrix whose rows correspond to layers and whose columns correspond to phonemes. The plotted matrices are hierarchically clustered using the same Euclidean-distance and average-linkage procedure as the saliency maps, allowing direct visual comparison of saliency reproducibility across splits. This analysis enables us to assess whether different architectures concentrate class-specific sensitivity in a small number of late modules or distribute it across multiple processing stages, and whether learned representations generalize between data standardization approaches.

%% file: nstc/results.tex
\section{Results and Discussion}
\label{sec:results_discussion}

\subsection{Benchmark Performance}
\label{subsec:benchmark_performance}

The prominent results are summarized in Table~\ref{tab:summary_results}, which reports F1-macro scores for various model configurations across the training, validation, test, and holdout splits. The table includes both our own ablation runs and external reference results from MEGConformer~\citep{dezuazo2025megconformer}. For the full set of configurations and metrics, see Table~\ref{tab:full_results} in the appendix.

Some configurations are not submitted to the holdout set, denoted by dashes. The PanPhon feature experiments omit training F1-macro because optimization happens on continuous articulatory vectors with an MSE loss. Producing phoneme predictions requires a nearest-neighbor projection back into the discrete inventory, which we only execute on validation and leaderboard splits to avoid repeatedly decoding the entire training corpus each epoch.

\begin{table}[h]
  \centering
  \scriptsize
  \setlength{\tabcolsep}{4pt}
  \renewcommand{\arraystretch}{0.8}
  \caption{F1-macro scores (\%) for LibriBrain phoneme classification.}
  \label{tab:summary_results}
  \begin{tabular}{lccccccl}
    \toprule
    \shortstack[c]{\textbf{Configuration}\\[-0.15ex]{\tiny \vphantom{(Ungrouped)}}} & \shortstack[c]{\textbf{Train}\\[-0.15ex]{\tiny \vphantom{(Ungrouped)}}} & \shortstack[c]{\textbf{Val.}\\[-0.15ex]{\tiny \vphantom{(Ungrouped)}}} & \shortstack[c]{\textbf{Test}\\[-0.15ex]{\tiny \vphantom{(Ungrouped)}}} & \shortstack[c]{\textbf{Test}\\[-0.15ex]{\tiny (Ungrouped)}} & \shortstack[c]{\textbf{Holdout}\\[-0.15ex]{\tiny \vphantom{(Ungrouped)}}} \\
    \midrule
    CNN, Label-Balancing, Repeat $\times 10$ & 90.81 & \textbf{71.95} & 47.47 & 2.73 & \textbf{35.40} \\
    CNN, +LayerNorm & 96.62 & 44.49 & 43.17 & -- & 24.40 \\
    CNN, No Group Averaging & \underline{34.55} & \underline{45.08} & \underline{39.53} & N/A & \underline{13.20} \\
    CNN + Augmentation (Section~\ref{subsubsec:augmentation}) & 48.55 & 49.31 & 34.03 & -- & 18.80  \\
    CNN-Transformer & 85.83 & 68.02 & 30.70 & 0.14 & 3.90  \\
    CNN, Label-Balancing, Repeat $\times 10$, +InstanceNorm & \textbf{99.43} & 62.75 & \textbf{60.95} & 4.00 & -- \\
    CNN--Transformer, +InstanceNorm & 83.60 & 57.72 & 55.87 & 3.83 & -- \\
    STFT CNN ($N_{\text{fft}}=25,H=5$) & 62.63 & 43.62 & 15.91 & 1.91 & -- \\
    STFT CNN ($N_{\text{fft}}=25,H=5$), +InstanceNorm & 86.54 & 36.89 & 25.90 & 1.95 & -- \\
    CNN--Transformer, Ternary PanPhon & -- & 51.68 & 3.33 & -- & -- \\
    CNN + Dist. Mapper & -- & -- & 0.06 & -- & --  \\
    \midrule
    \citep{dezuazo2025megconformer} (10 epochs) & 61.10 & 51.55 & 49.99 & 3.42 & -- \\
    \citep{dezuazo2025megconformer} (36 epochs) & 68.55 & 65.67 & \textbf{64.09} & 3.88 & -- \\
    \bottomrule
  \end{tabular}
\end{table}

\subsubsection{Main Findings}

Although validation and test contain the same raw samples, they are standardized with different statistics, so the validation--test gap directly probes robustness to preprocessing and distribution shift. Indeed, across runs with holdout scores, the gap between test and holdout performance is consistent at around 20 points, which suggests that the test split is a reasonable proxy for holdout performance when holdout submissions are unavailable. Overall, we treat the holdout split as the main benchmark, following the official evaluation metric, but the test split is also informative for understanding robustness to preprocessing shift. 

Among our own models, the strongest test result is obtained by the label-balanced CNN with repeated group averaging and instance normalization, which reaches 60.95\% on the test split. This is a gain of +13.48 points over the corresponding non-InstanceNorm model at 47.47\%, and it narrows the gap to the external MEGConformer reference, which reaches 64.09\% after a longer training schedule. The same pattern also appears in the CNN--Transformer family, where adding InstanceNorm raises test performance from 4.29\% to 55.87\%, indicating that a large share of the observed generalizability is tied to normalization rather than architecture alone.

\subsubsection{Ablation Study}
\label{subsec:ablation_study}

Table~\ref{tab:full_results} in the appendix isolates the main design choices. The following paragraphs summarize the most relevant trends.

\paragraph{Normalization layers.}
InstanceNorm is the most effective single modification observed in our experiments. For the strongest CNN setting, adding InstanceNorm raises test performance from 47.47\% to 60.95\% (+13.48). The same trend is even more pronounced for the CNN--Transformer family, where InstanceNorm increases test performance from 4.29\% to 55.87\% (+51.58). By contrast, LayerNorm and BatchNorm mainly improve training fit without comparable transfer gains: relative to the repeat-1 CNN, LayerNorm raises training F1 from 80.06\% to 96.62\% but reduces validation from 71.77\% to 44.49\%, while BatchNorm reaches 91.50\% on training yet only 42.84\% on test. Taken together, these results suggest that instance normalization is better matched to the split shift in this task than statistics accumulated from the training set.

\paragraph{Group averaging and label balancing.}
Group averaging with label balancing provides the largest improvement among the repeat-1 CNN settings. Relative to the plain baseline, it raises test performance from 39.53\% to 44.28\% (+4.75) and holdout performance from 13.20\% to 21.60\% (+8.40). We therefore treat group averaging and label balancing as the main non-normalization gains, while attributing the additional improvement of the best CNN separately to repeated grouping.

\paragraph{Repeated grouping.}
Repeated grouping provides an additional robustness gain. For the same CNN backbone, increasing the repeat count from 1 to 10 changes validation from 71.77\% to 71.95\% (+0.18), test from 44.28\% to 47.47\% (+3.19), and holdout from 21.60\% to 35.40\% (+13.80). The effect is therefore modest on validation but substantial on holdout.

\paragraph{Architecture and input representation.}
Architectural changes alone are not sufficient. Without InstanceNorm, the CNN--Transformer variants remain unstable, reaching only 4.29\% or 0.98\% on test despite validation scores above 63\%. The STFT CNN variants are also consistently weak, with the best test result at 25.90\%, far below the strongest time-domain CNN. Data augmentation shows mixed effects: one augmented CNN improves test from 39.53\% to 47.74\%, but its holdout score collapses to 3.60\%, while the simpler augmented setting reaches only 34.03\% on test and 18.80\% on holdout. These runs reinforce that raw test gains are not sufficient unless they remain consistent on holdout.

\paragraph{Alternative label spaces.}
Alternative label spaces also fail to transfer. The PanPhon variants reach 51.68\% and 57.21\% on validation, but their test scores fall to 3.33\% and 0.88\%, while the distribution-mapper variants are near zero. In our setup, changing the label space therefore does not solve the shift problem.

Overall, our ablation study suggests a two-part conclusion: group averaging, label balancing, and repeated grouping produce the strongest verified gains on holdout, while InstanceNorm is the most promising route for improving generalizability.

\subsection{Saliency Map Analysis}
\label{subsec:saliency_map_analysis}

\begin{figure}[h]
  \centering
  \captionsetup{type=figure}
  \captionsetup[subfigure]{justification=centering}

  \begin{subfigure}[t]{0.49\linewidth}
    \centering
    \includegraphics[width=\linewidth]{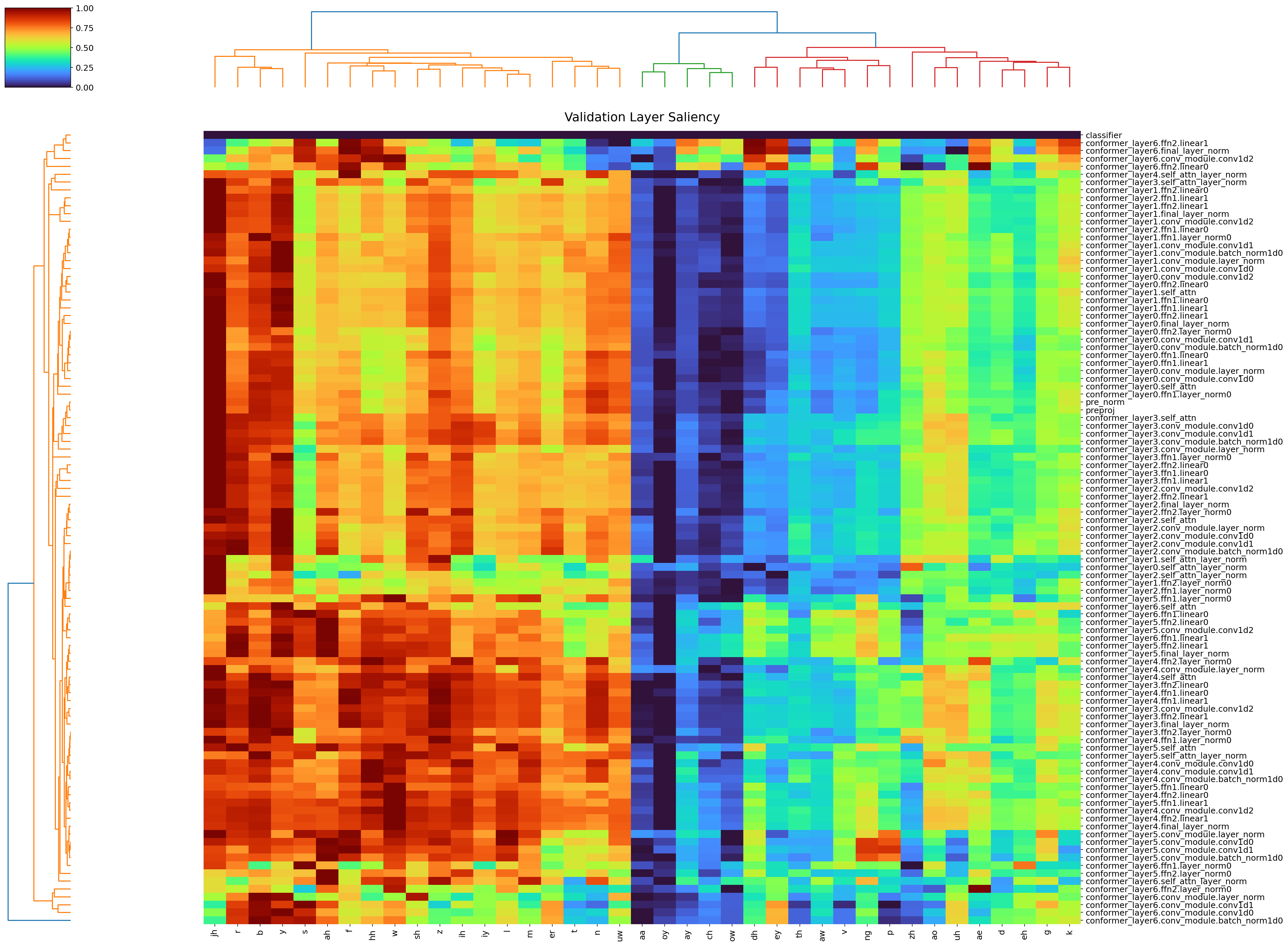}
    \caption{Validation}
  \end{subfigure}\hfill
  \begin{subfigure}[t]{0.49\linewidth}
    \centering
    \includegraphics[width=\linewidth]{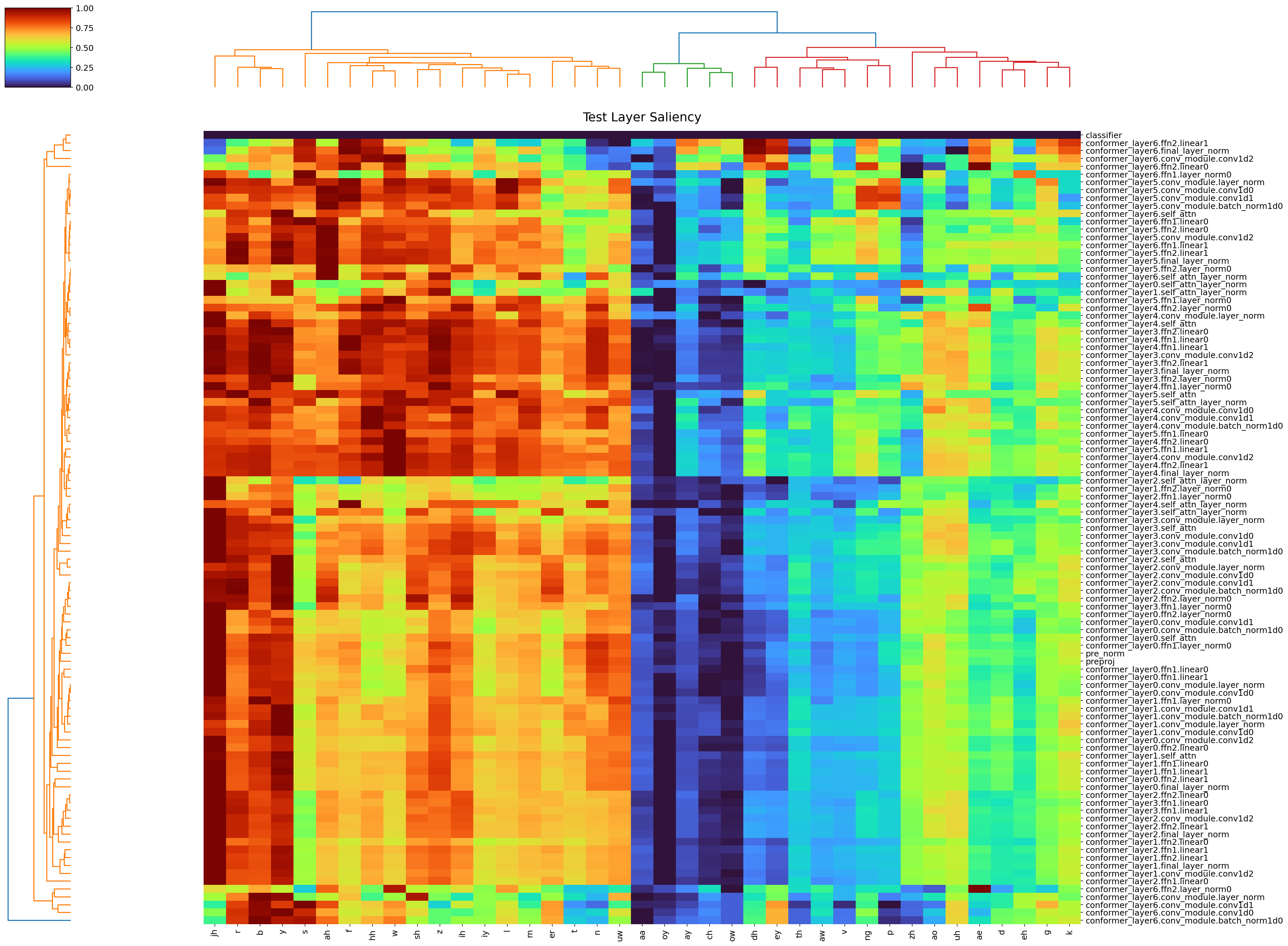}
    \caption{Test}
  \end{subfigure}

  \caption{Row-wise normalized layer saliency maps for MEGConformer across validation and test splits. The saliency is visually stable across splits and distributed across encoder blocks, matching the strong validation--test transfer reported in Table~\ref{tab:summary_results}.}
  \label{fig:saliency_map_conformer}
\end{figure}

\begin{figure}[t]
  \centering
  \captionsetup{type=figure}
  \captionsetup[subfigure]{justification=centering}

  \textbf{Without InstanceNorm}\par\smallskip
  \textit{Validation}\par\medskip
  \begin{subfigure}[t]{0.32\linewidth}
    \centering
    \includegraphics[width=\linewidth]{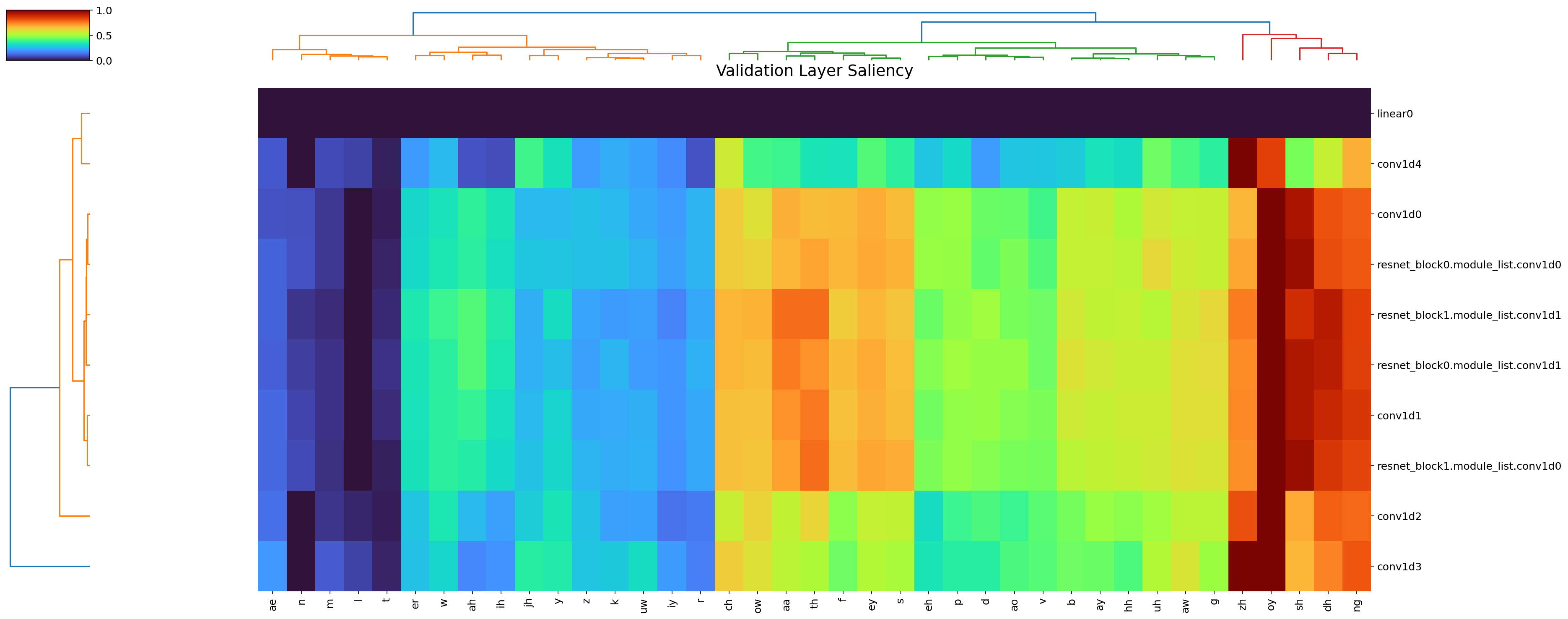}
    \caption{ResNet CNN}
  \end{subfigure}\hfill
  \begin{subfigure}[t]{0.32\linewidth}
    \centering
    \includegraphics[width=\linewidth]{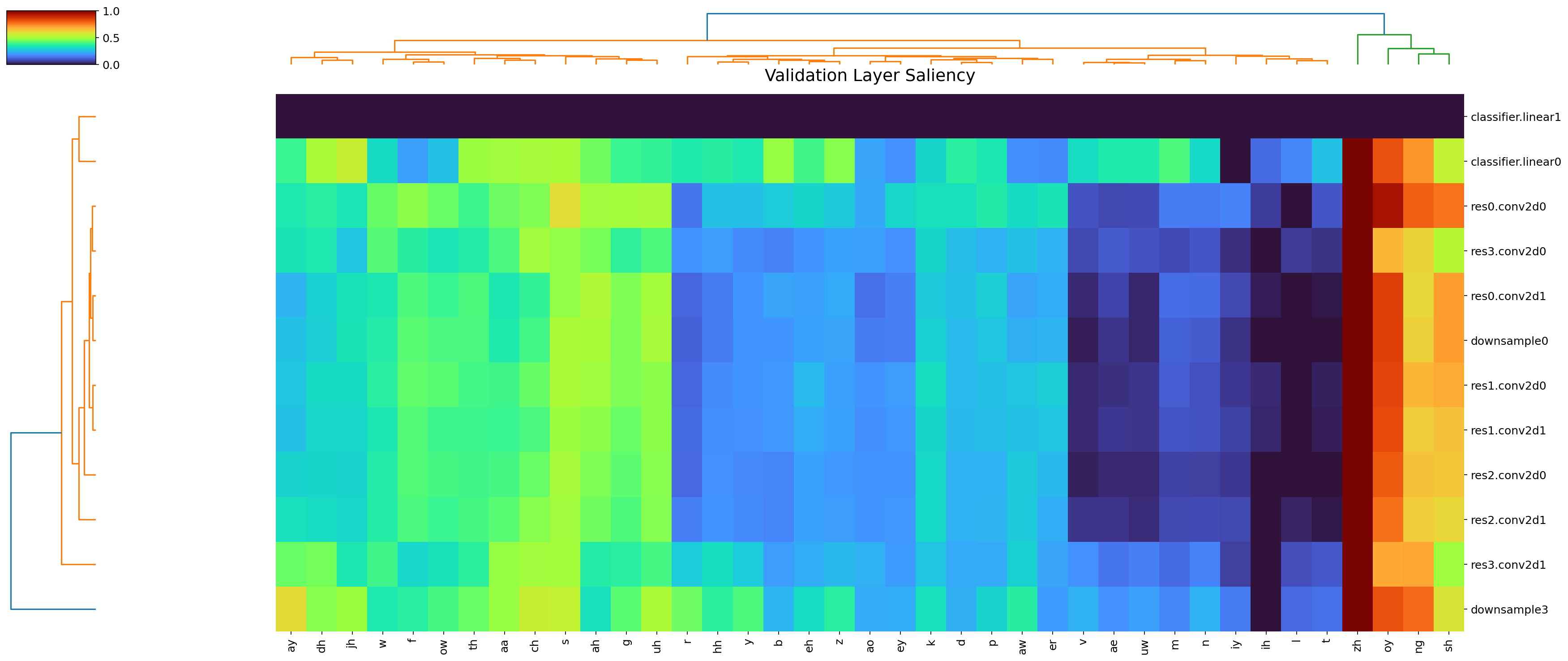}
    \caption{STFT CNN}
  \end{subfigure}\hfill
  \begin{subfigure}[t]{0.32\linewidth}
    \centering
    \includegraphics[width=\linewidth]{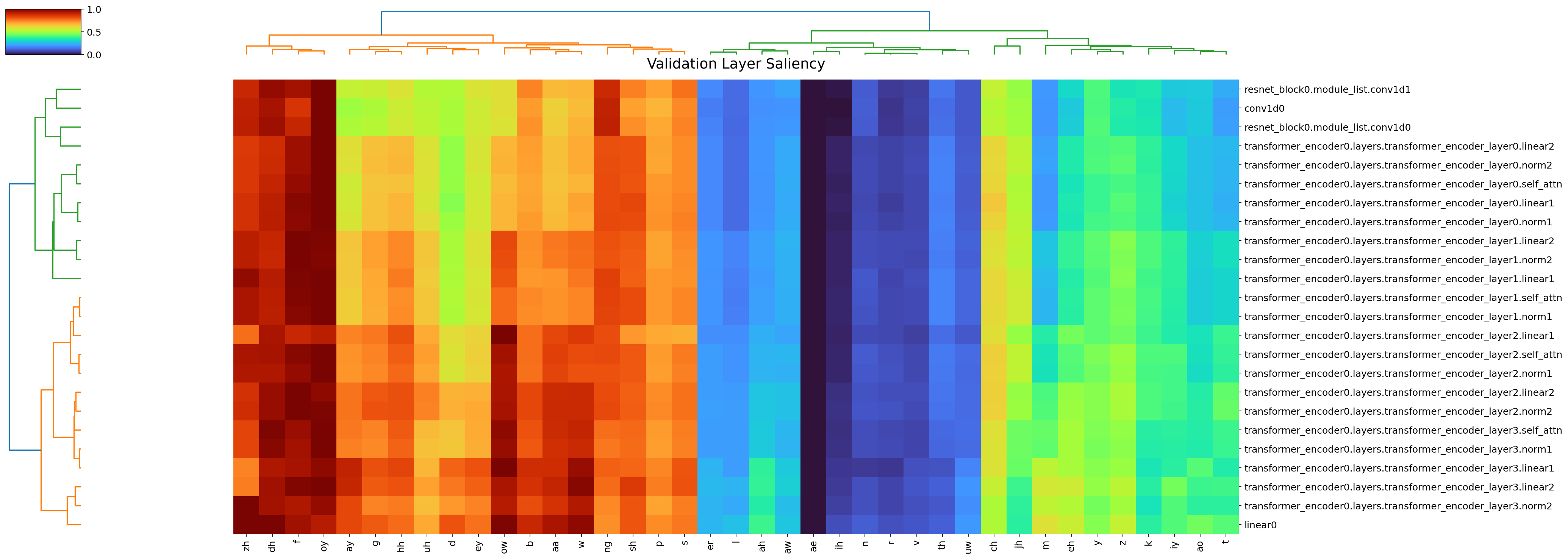}
    \caption{CNN--Transformer}
  \end{subfigure}

  \medskip

  \textit{Test}\par\medskip
  \begin{subfigure}[t]{0.32\linewidth}
    \centering
    \includegraphics[width=\linewidth]{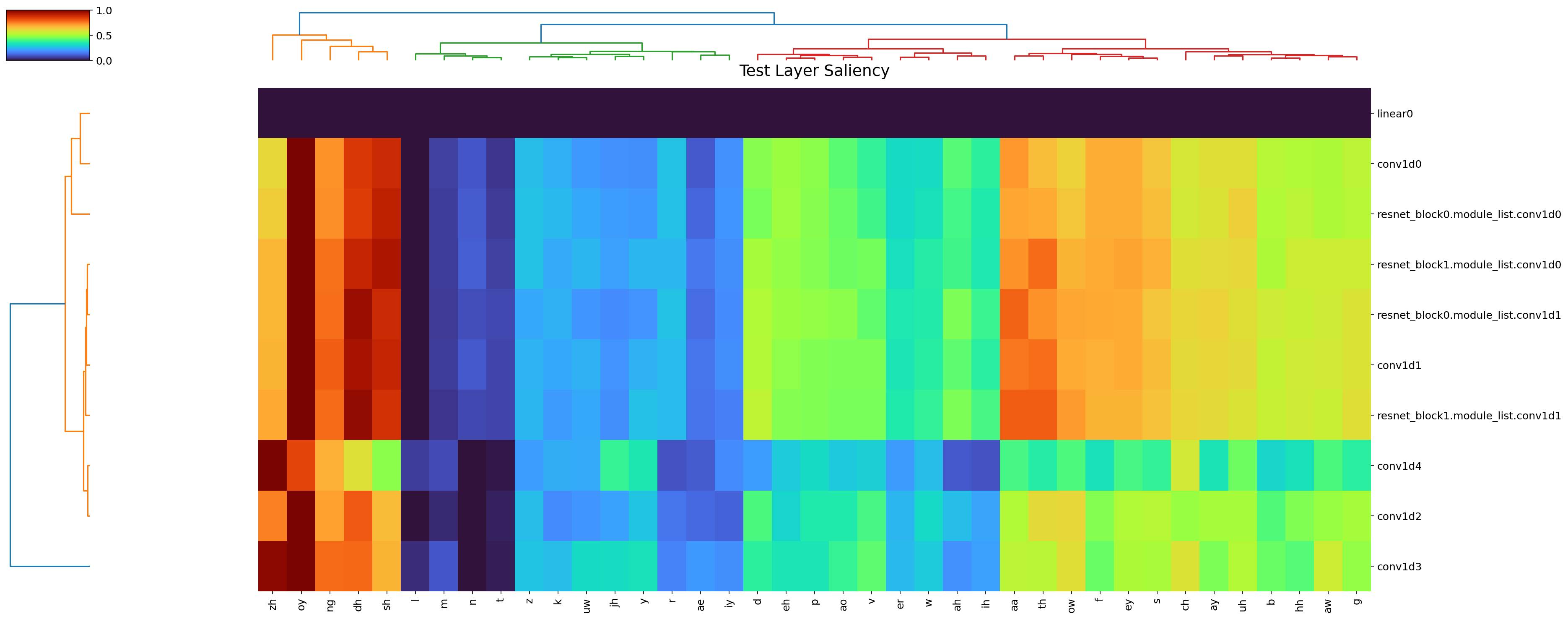}
    \caption{ResNet CNN}
  \end{subfigure}\hfill
  \begin{subfigure}[t]{0.32\linewidth}
    \centering
    \includegraphics[width=\linewidth]{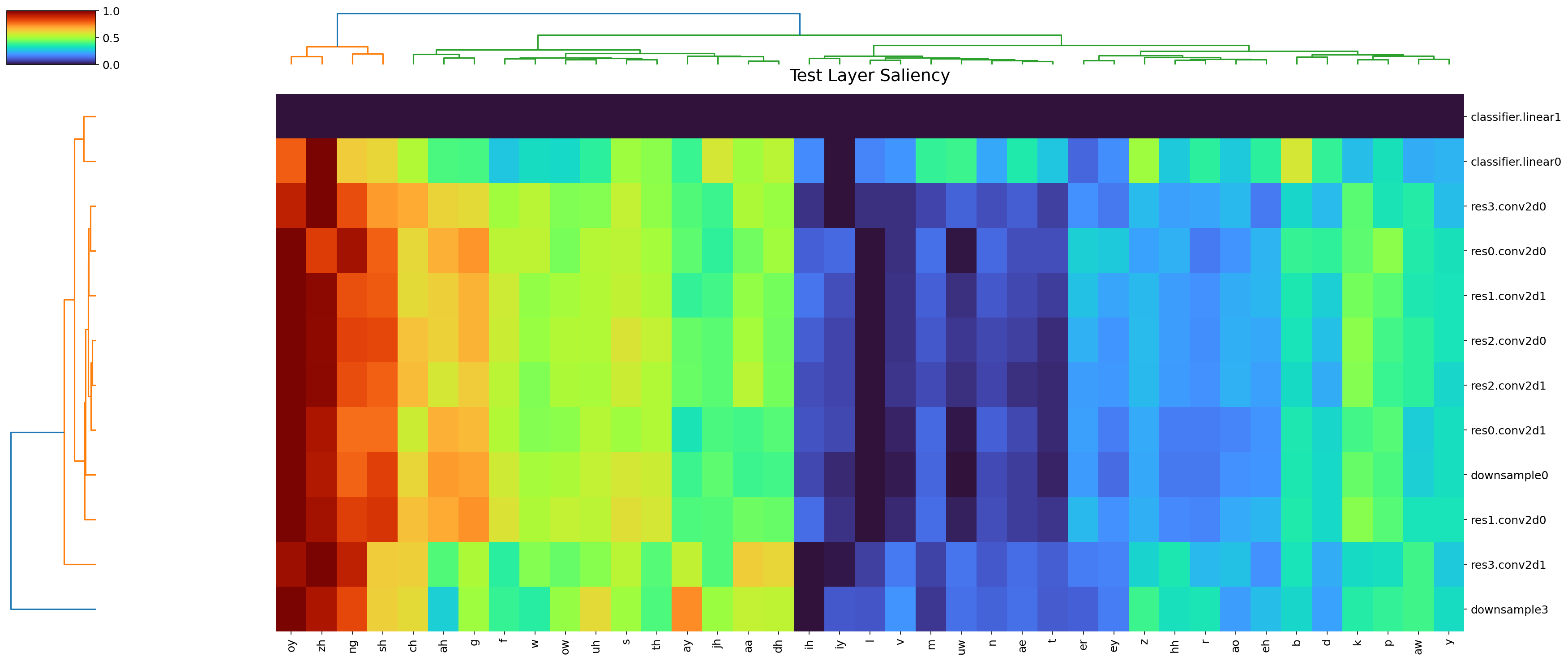}
    \caption{STFT CNN}
  \end{subfigure}\hfill
  \begin{subfigure}[t]{0.32\linewidth}
    \centering
    \includegraphics[width=\linewidth]{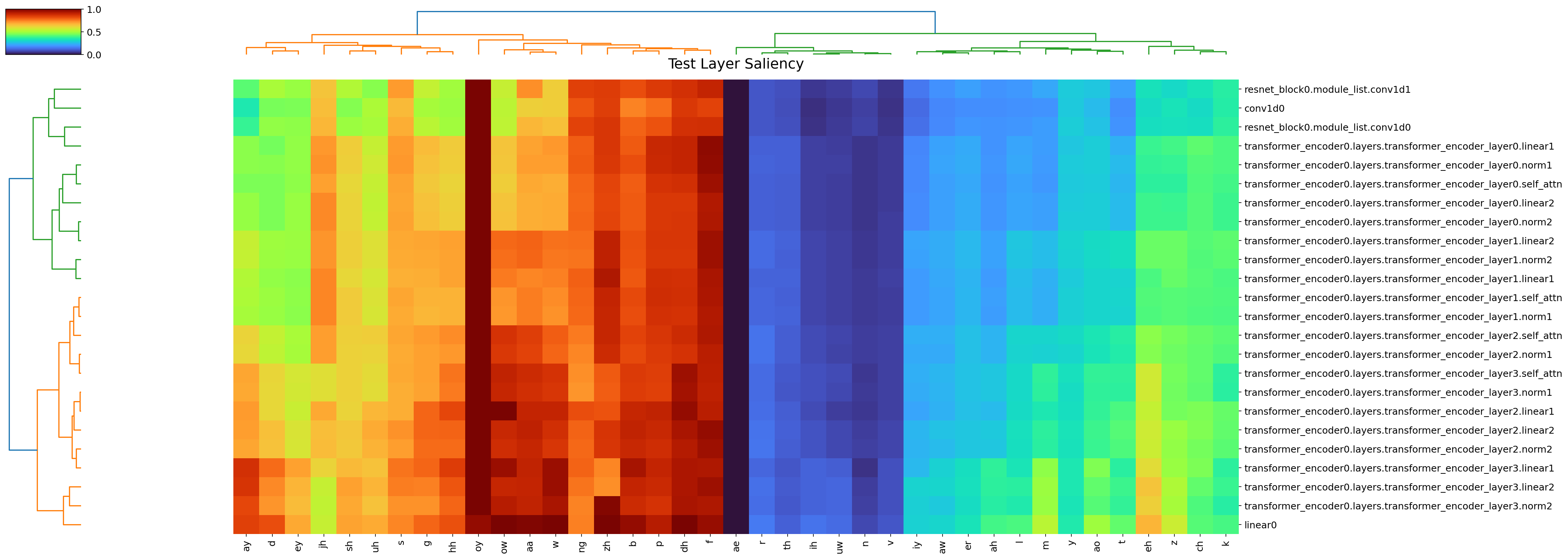}
    \caption{CNN--Transformer}
  \end{subfigure}

  \vspace{0.5em}

  \textbf{With InstanceNorm}\par\smallskip
  \textit{Validation}\par\medskip
  \begin{subfigure}[t]{0.32\linewidth}
    \centering
    \includegraphics[width=\linewidth]{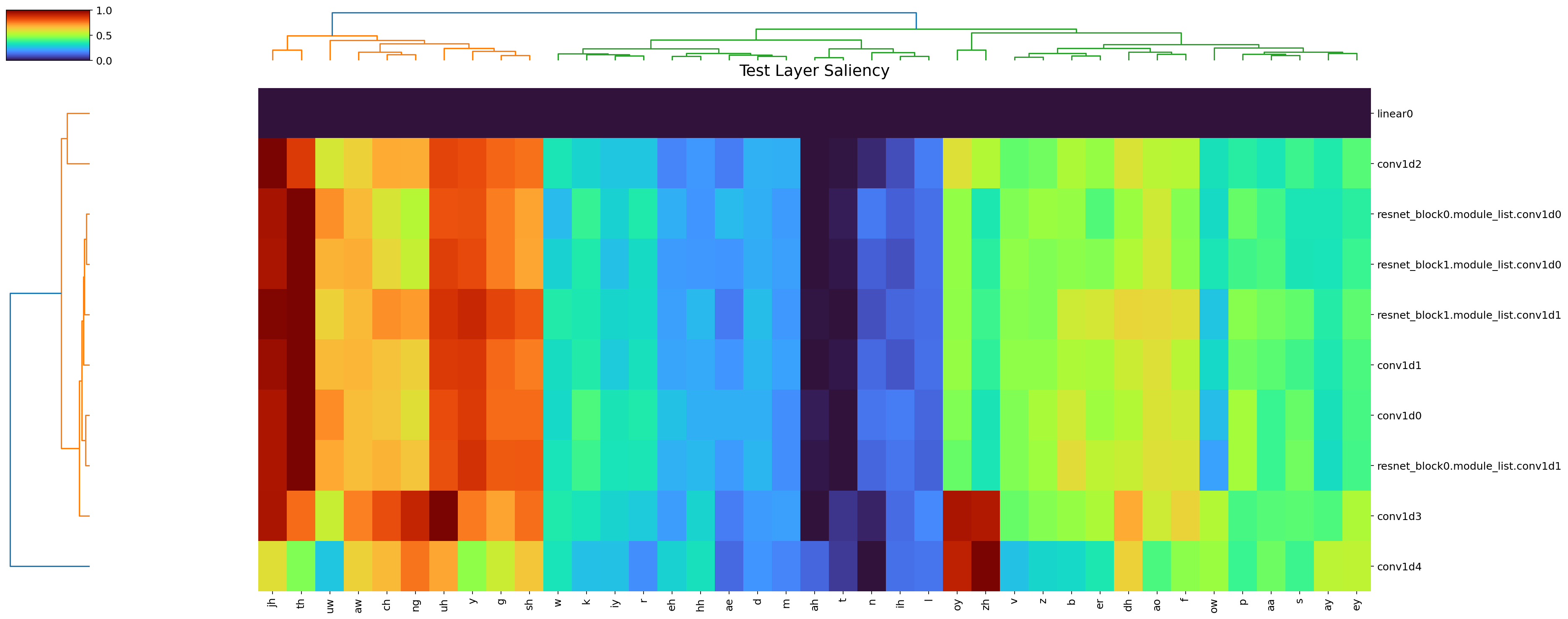}
    \caption{ResNet CNN + IN}
  \end{subfigure}\hfill
  \begin{subfigure}[t]{0.32\linewidth}
    \centering
    \includegraphics[width=\linewidth]{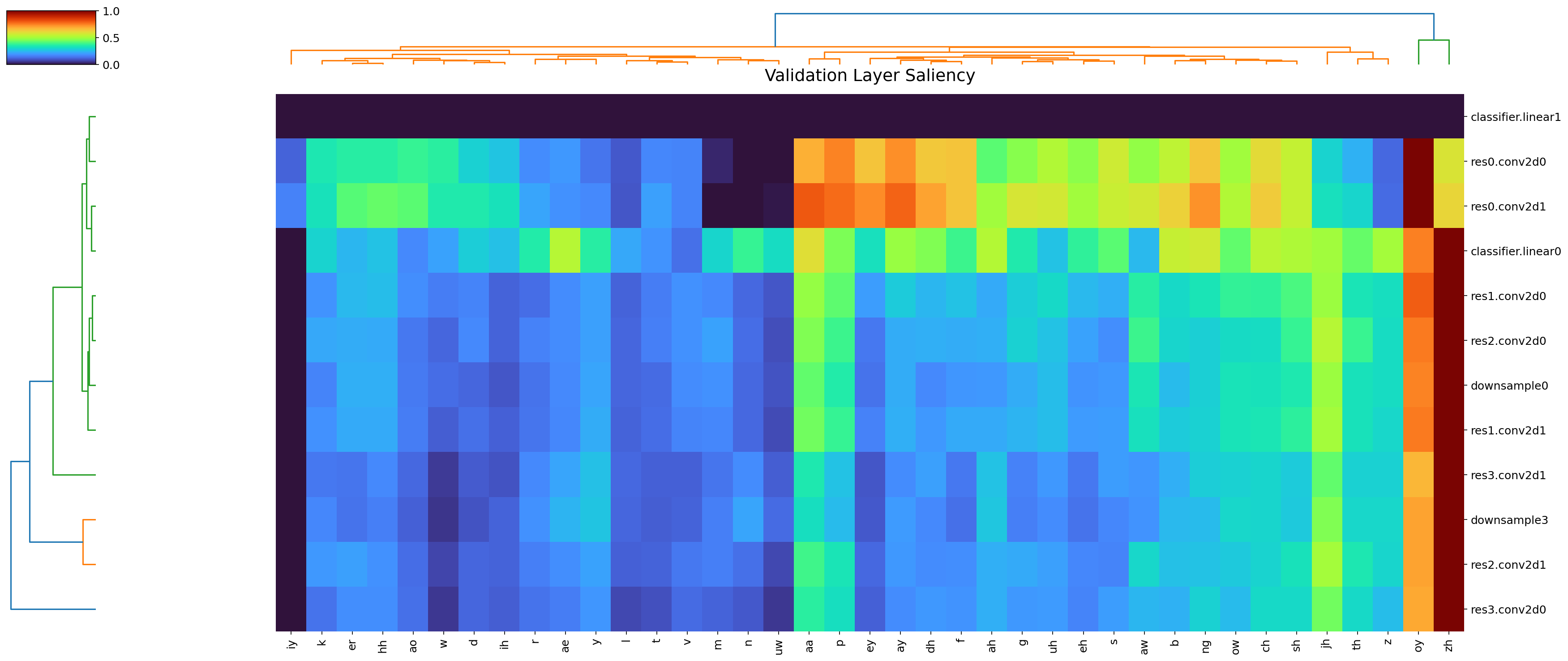}
    \caption{STFT CNN + IN}
  \end{subfigure}\hfill
  \begin{subfigure}[t]{0.32\linewidth}
    \centering
    \includegraphics[width=\linewidth]{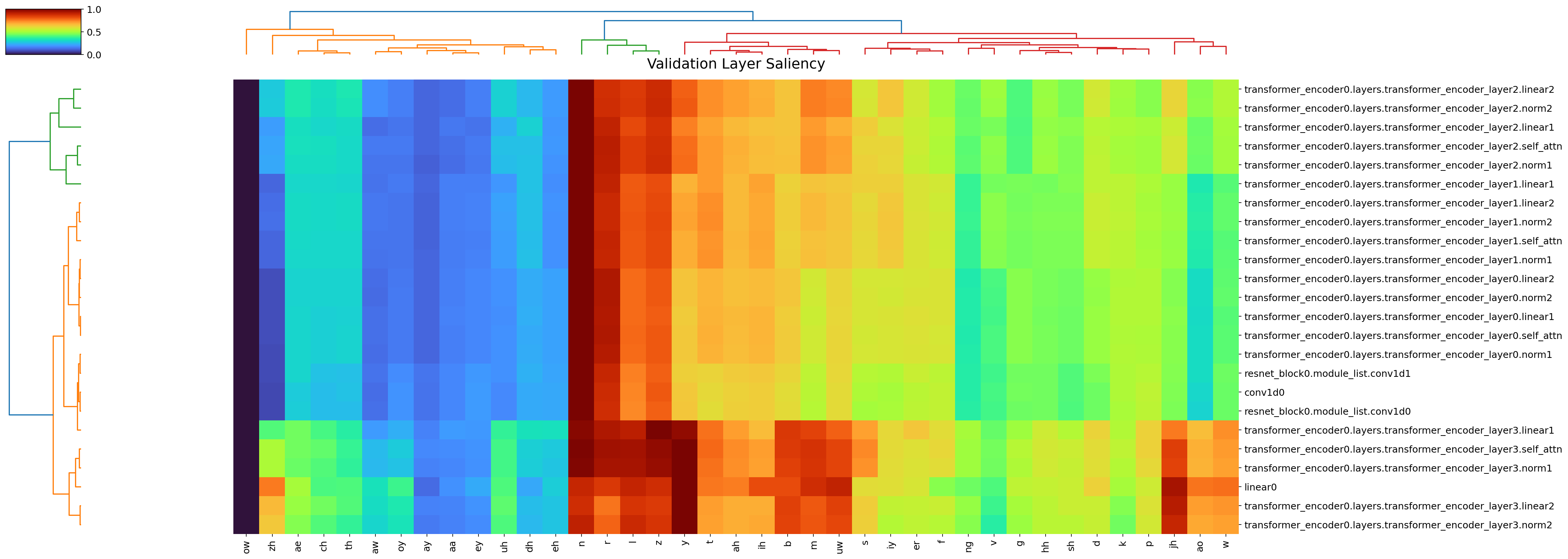}
    \caption{CNN--Transformer + IN}
  \end{subfigure}

  \medskip

  \textit{Test}\par\medskip
  \begin{subfigure}[t]{0.32\linewidth}
    \centering
    \includegraphics[width=\linewidth]{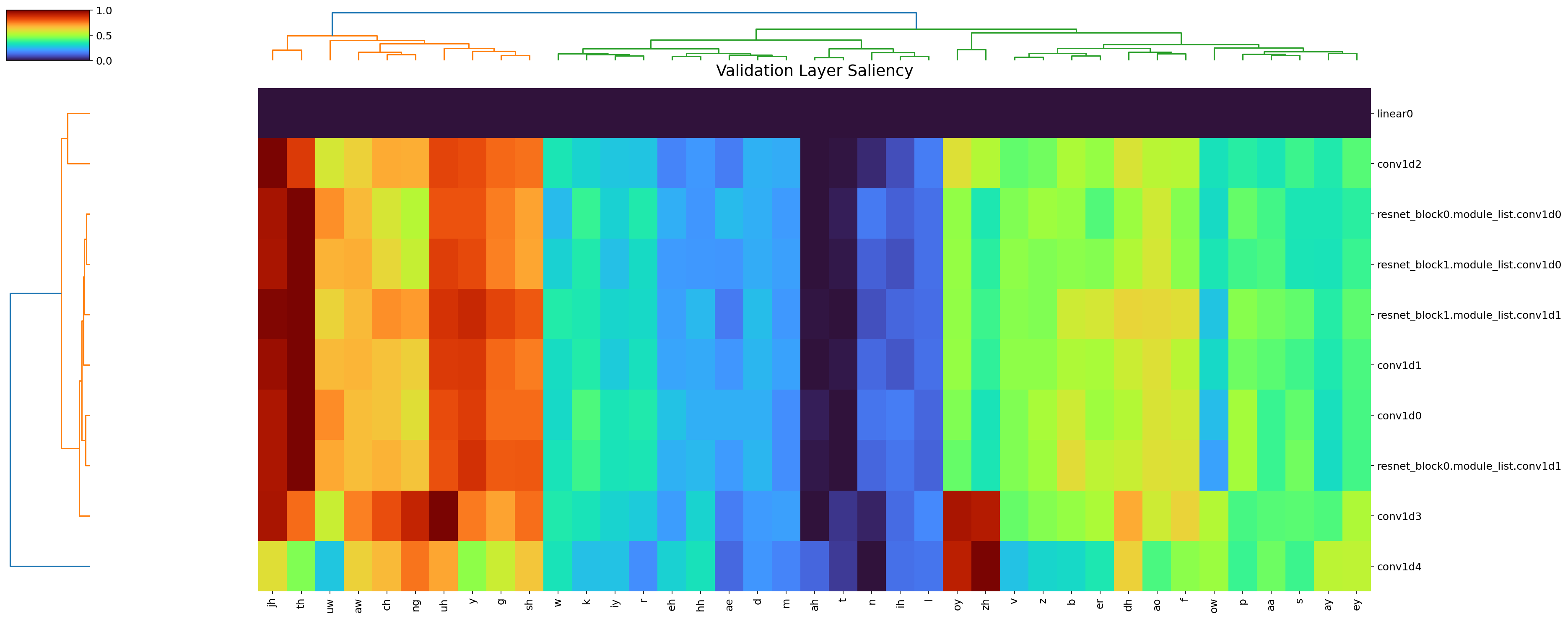}
    \caption{ResNet CNN + IN}
  \end{subfigure}\hfill
  \begin{subfigure}[t]{0.32\linewidth}
    \centering
    \includegraphics[width=\linewidth]{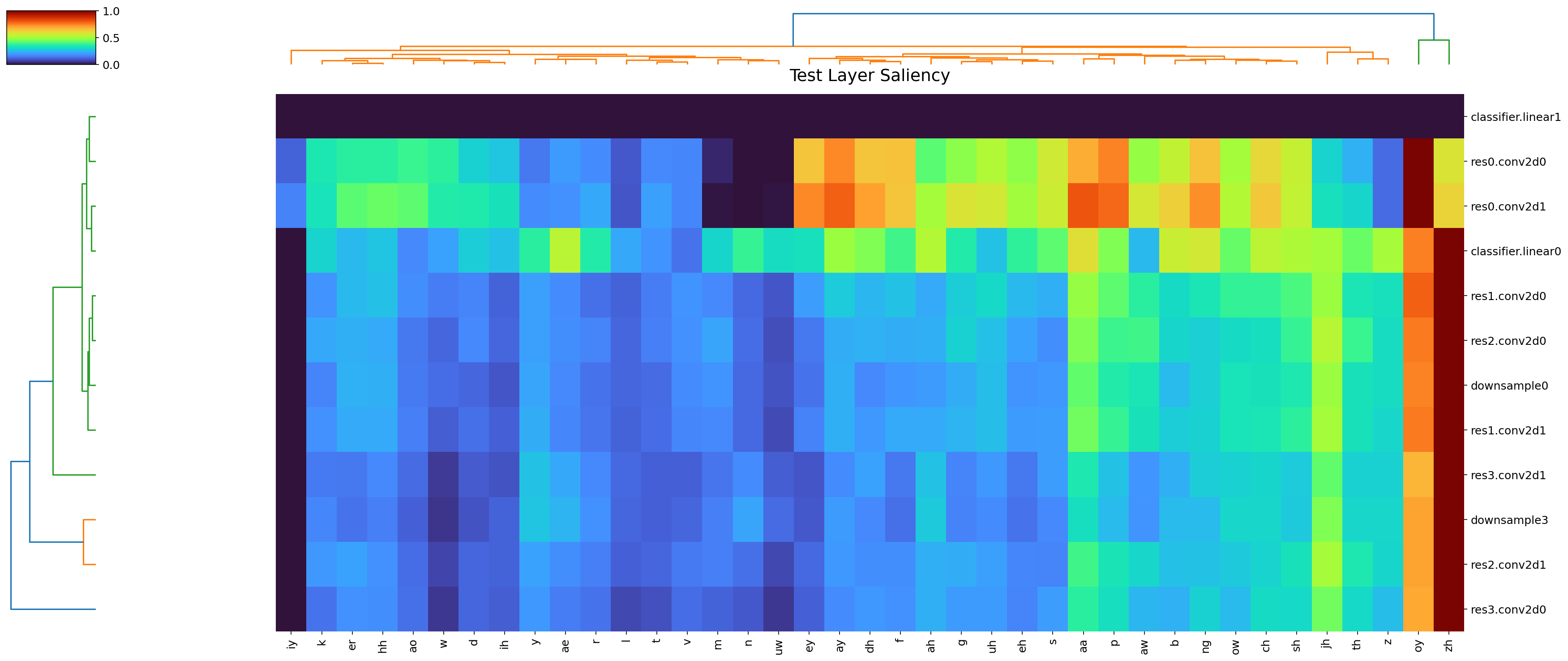}
    \caption{STFT CNN + IN}
  \end{subfigure}\hfill
  \begin{subfigure}[t]{0.32\linewidth}
    \centering
    \includegraphics[width=\linewidth]{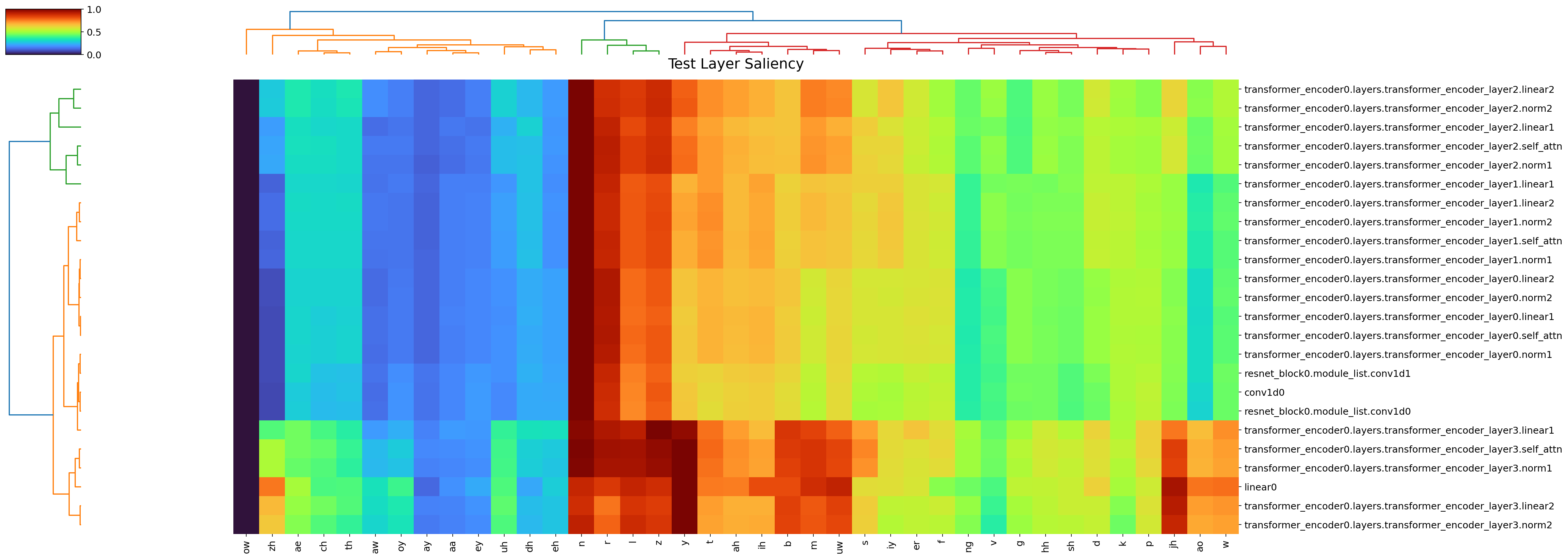}
    \caption{CNN--Transformer + IN}
  \end{subfigure}

  \caption{Row-wise normalized layer saliency maps for the three models for which both standard and InstanceNorm variants are available. In each 3-column block, the upper row shows validation maps and the lower row shows test maps. Each row within a saliency map corresponds to a trainable sublayer, and each column corresponds to a phoneme class. The color intensity indicates the relative saliency of each phoneme for that layer, normalized within the row.}
  \label{fig:saliency_map_grid}
\end{figure}

Figures~\ref{fig:saliency_map_conformer} and~\ref{fig:saliency_map_grid} compare row-wise normalized layer saliency maps, where each row corresponds to a trainable sublayer and each column to a phoneme. Because the values are normalized within each row, the colors should be read as showing which phonemes each layer is relatively most sensitive to, rather than comparing absolute gradient magnitude across different layers or models.

\subsubsection{Phoneme-wise patterns}

Across Figures~\ref{fig:saliency_map_grid} and~\ref{fig:saliency_map_conformer}, the phoneme axis is far from uniform. Most models partition the phonemes into three broad groups: a compact high-saliency cluster, a compact low-saliency cluster, and a larger intermediate group with smoother gradients across layers. The high-saliency cluster is repeatedly anchored by a small set of marked or relatively infrequent phones, especially \textit{zh}, \textit{oy}, \textit{sh}, \textit{dh}, and often \textit{ng}; the low-saliency side is more often associated with common sonorants and simple vowels such as \textit{l}, \textit{n}, \textit{m}, \textit{ae}, and \textit{iy}. The main difference between models is therefore not whether phoneme grouping exists, but whether the same grouping is preserved across splits and distributed across many layers rather than collapsing onto a few isolated columns.

For the \textbf{ResNet CNN} without InstanceNorm, the phoneme grouping is unstable: the validation and test maps both highlight a narrow cluster around phones such as \textit{zh}, \textit{oy}, \textit{sh}, \textit{dh}, and \textit{ng}, but this cluster changes position and its contrast against the remaining phonemes shifts noticeably across splits. Adding InstanceNorm makes the same backbone much more structured. The ResNet CNN + InstanceNorm maps show an almost unchanged three-block partition between validation and test, with a compact high-saliency group, a clearly suppressed group centered on sonorants and alveolar consonants, and a broader middle group occupying the rest of the inventory.

The \textbf{STFT CNN} exhibits the strongest concentration. Without InstanceNorm, much of the map is driven by a very small subset of columns, again dominated by phones such as \textit{zh}, \textit{oy}, \textit{ng}, and \textit{sh}, while many vowels and sonorants remain weak over nearly all rows. This makes the grouping visually coarse and split-dependent. InstanceNorm stabilizes the pattern and restores a clearer low-saliency block versus high-saliency block, but the STFT model still appears more peaked and less evenly distributed than the other normalized models.

\textbf{CNN--Transformer} lies between these extremes. Even without InstanceNorm, it already shows recognizable phoneme clusters rather than a purely noisy map, but its grouping is still uneven: a hot block is concentrated on a subset of consonants, a cold block remains centered on alveolar consonants like \textit{l}, \textit{n}, and several simple vowels, and some late layers form their own local pattern instead of following the dominant partition. With InstanceNorm, these local deviations become much smaller and the validation/test maps become nearly identical, yielding a stable three-group structure similar to the normalized CNN.

\textbf{MEGConformer} displays the clearest overall organization. Its validation and test maps preserve almost the same phoneme dendrogram and nearly the same layer-wise color pattern, with one large high-saliency group, one narrow low-saliency group, and one intermediate group that remain consistent across encoder blocks. Compared with our models, the important qualitative difference is that the phoneme grouping in MEGConformer is both more stable across splits and more broadly shared across layers, which matches its stronger generalization performance.

\subsubsection{Cross-split stability}

\begin{figure}[h]
  \centering
  \captionsetup{type=figure}
  \captionsetup[subfigure]{justification=centering}

  \begin{subfigure}[t]{0.49\linewidth}
    \centering
    \includegraphics[width=\linewidth]{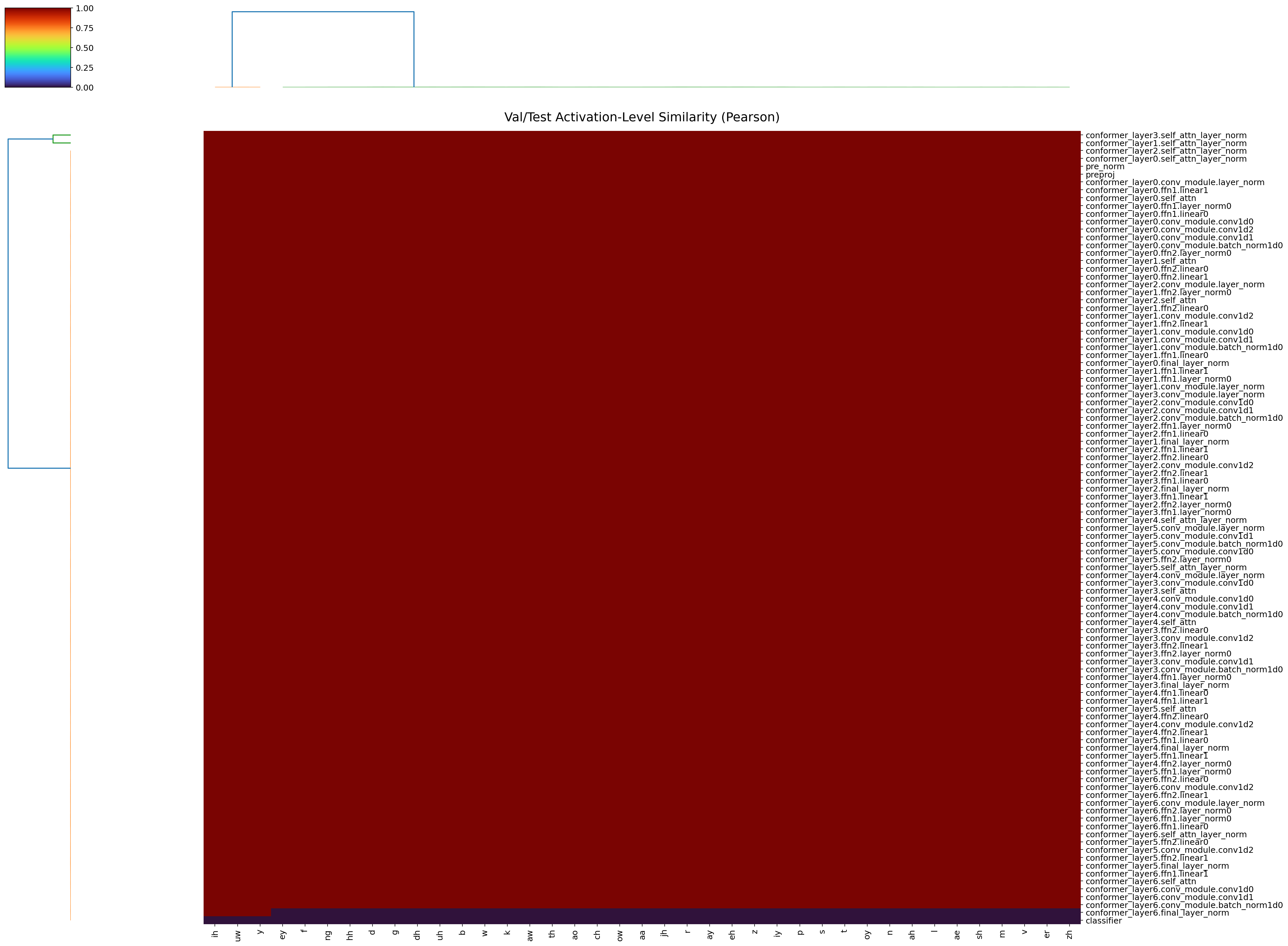}
    \caption{Pearson}
  \end{subfigure}\hfill
  \begin{subfigure}[t]{0.49\linewidth}
    \centering
    \includegraphics[width=\linewidth]{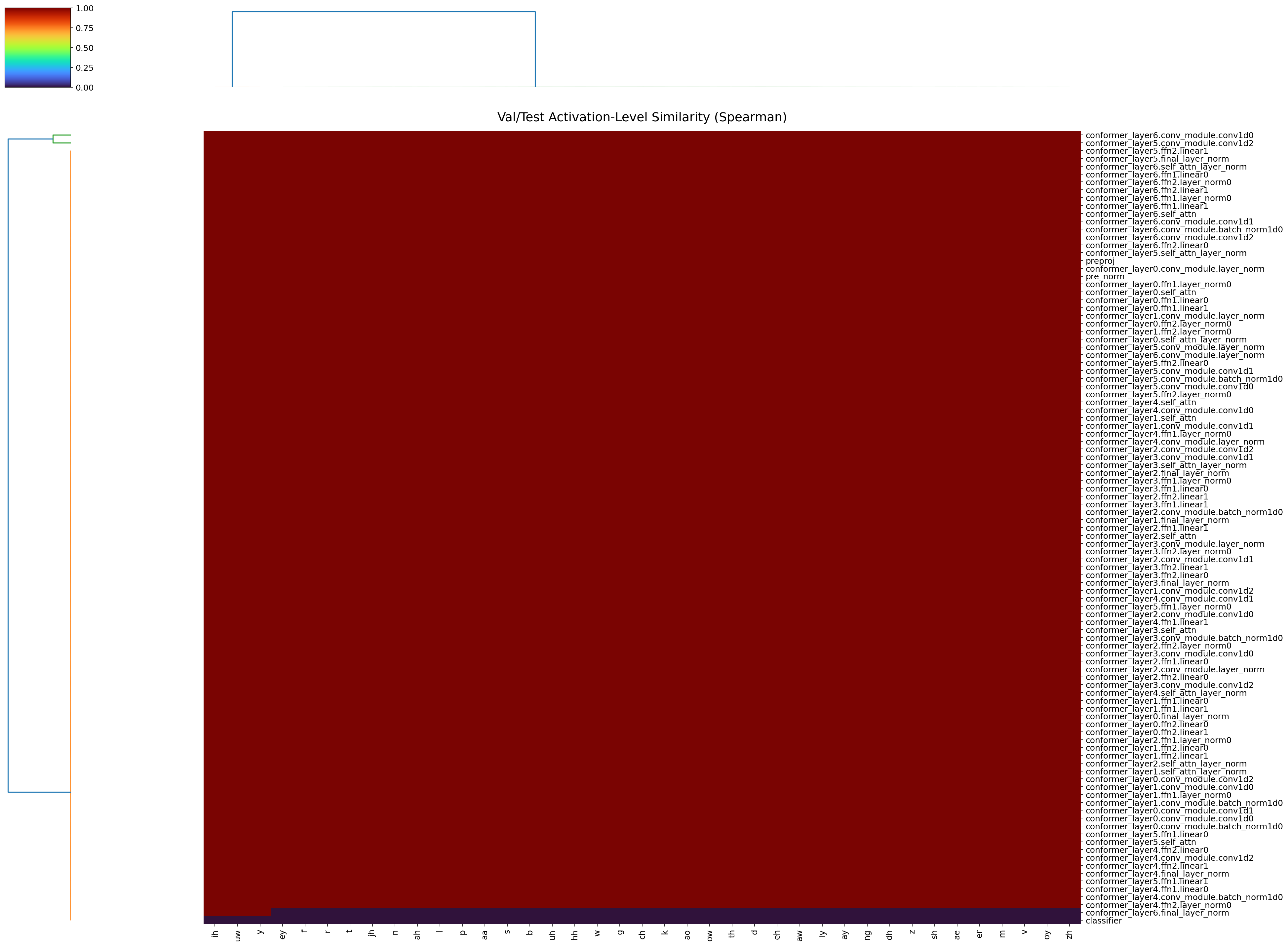}
    \caption{Spearman}
  \end{subfigure}

  \caption{Layer-by-phoneme validation-test saliency similarity matrices for MEGConformer. Both metrics show near-uniformly high correspondence across layers and phoneme classes.}
  \label{fig:saliency_similarity_conformer}
\end{figure}

\begin{figure}[h]
  \centering
  \captionsetup{type=figure}
  \captionsetup[subfigure]{justification=centering}

  \textbf{Without InstanceNorm}\par\smallskip
  \textit{Pearson}\par\medskip
  \begin{subfigure}[t]{0.32\linewidth}
    \centering
    \includegraphics[width=\linewidth]{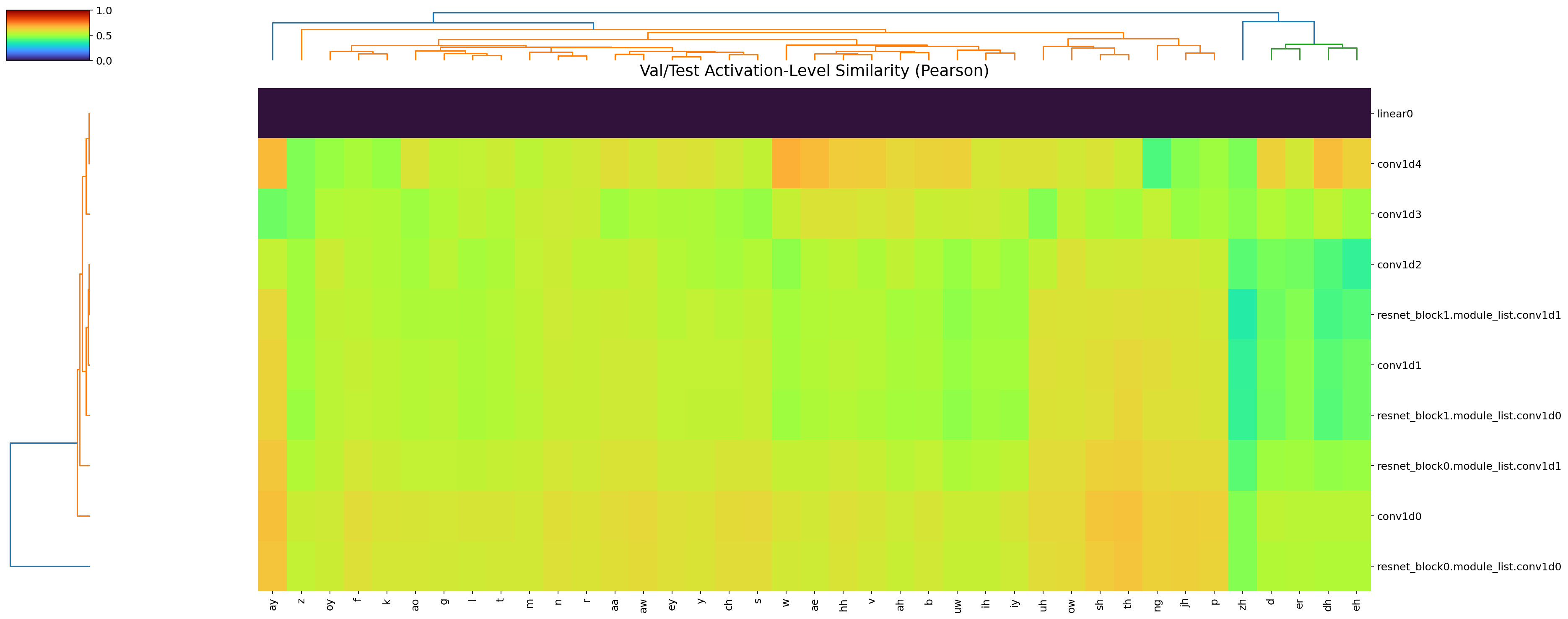}
    \caption{ResNet CNN}
  \end{subfigure}\hfill
  \begin{subfigure}[t]{0.32\linewidth}
    \centering
    \includegraphics[width=\linewidth]{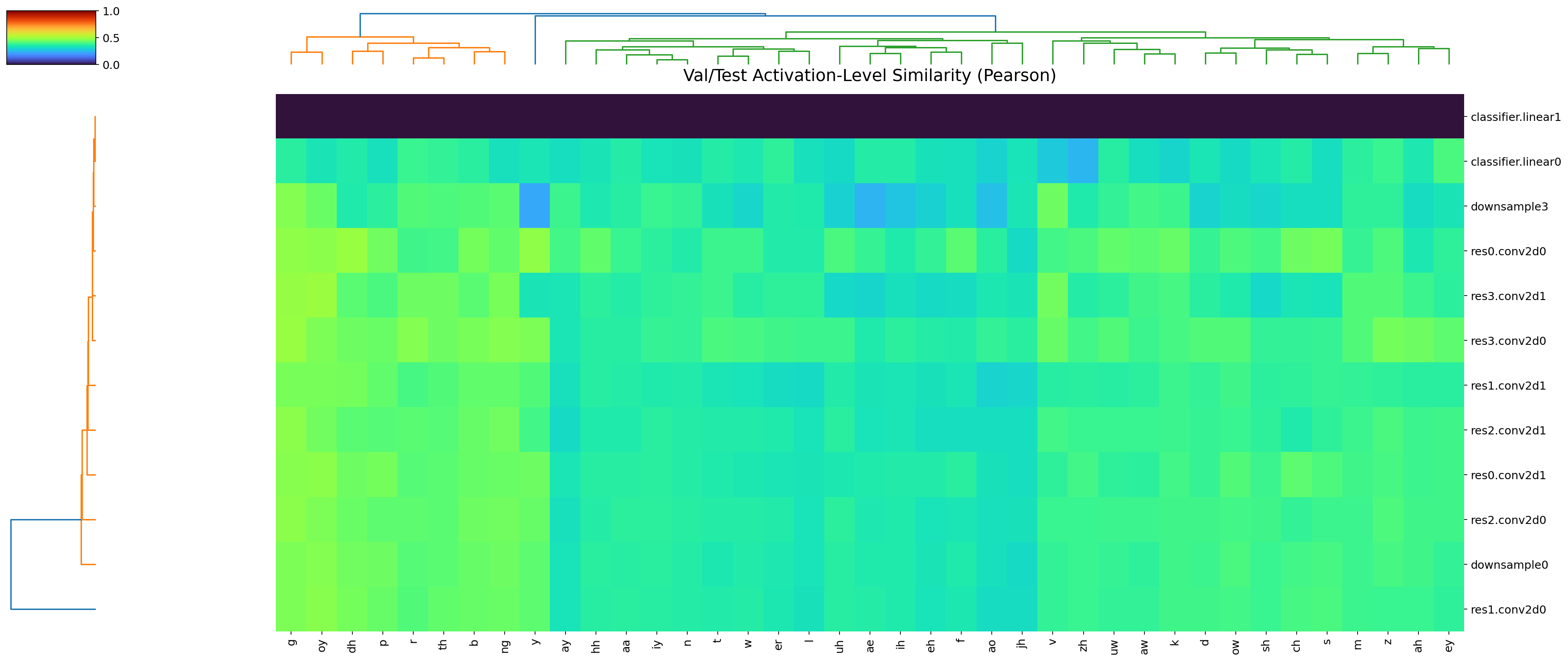}
    \caption{STFT CNN}
  \end{subfigure}\hfill
  \begin{subfigure}[t]{0.32\linewidth}
    \centering
    \includegraphics[width=\linewidth]{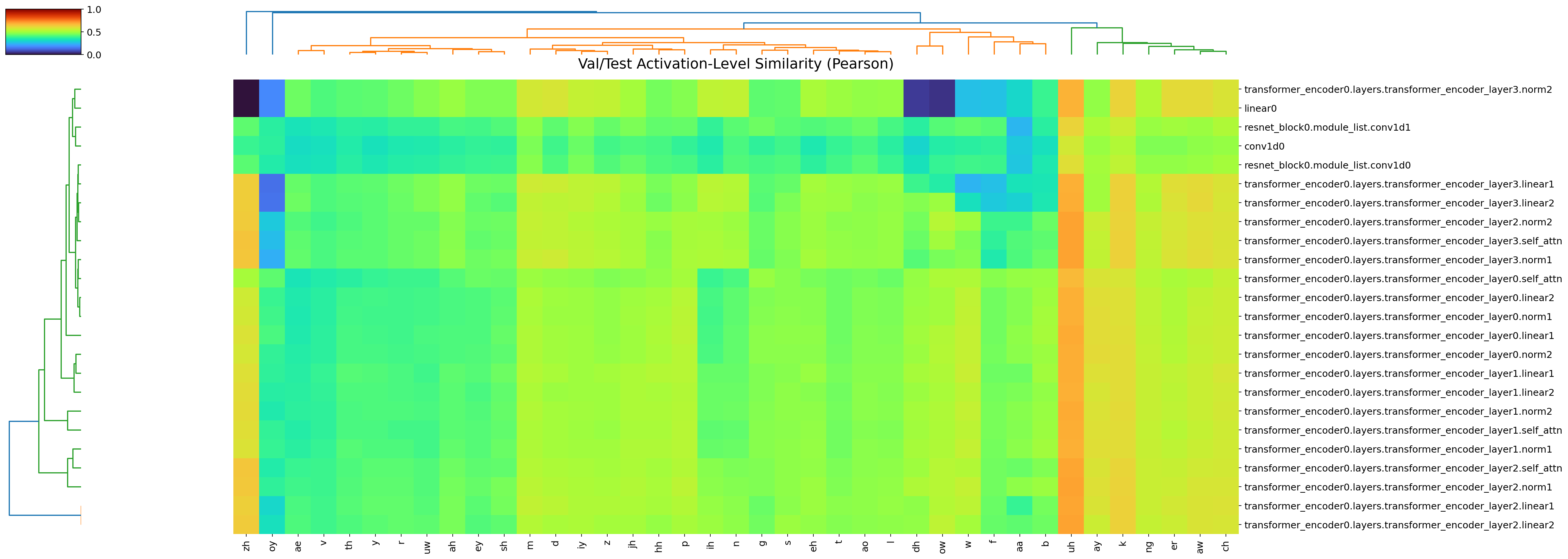}
    \caption{CNN--Transformer}
  \end{subfigure}

  \medskip

  \textit{Spearman}\par\medskip
  \begin{subfigure}[t]{0.32\linewidth}
    \centering
    \includegraphics[width=\linewidth]{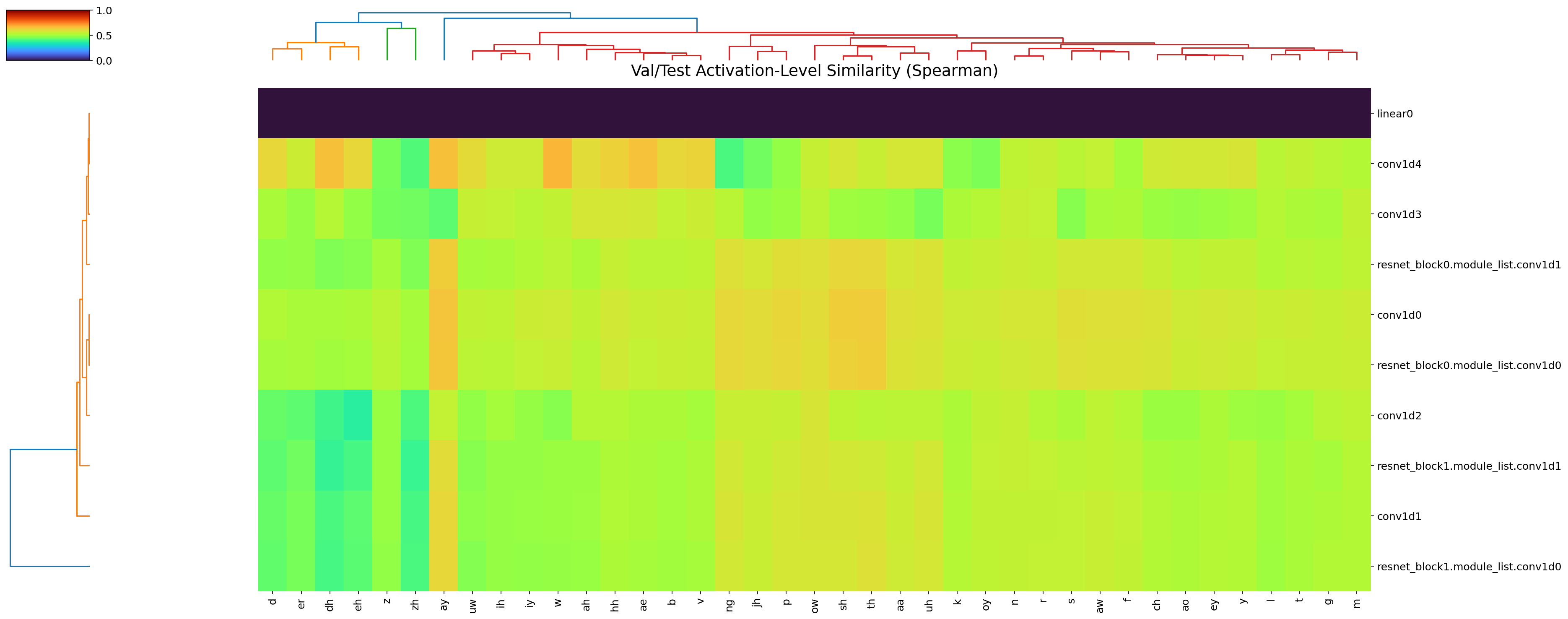}
    \caption{ResNet CNN}
  \end{subfigure}\hfill
  \begin{subfigure}[t]{0.32\linewidth}
    \centering
    \includegraphics[width=\linewidth]{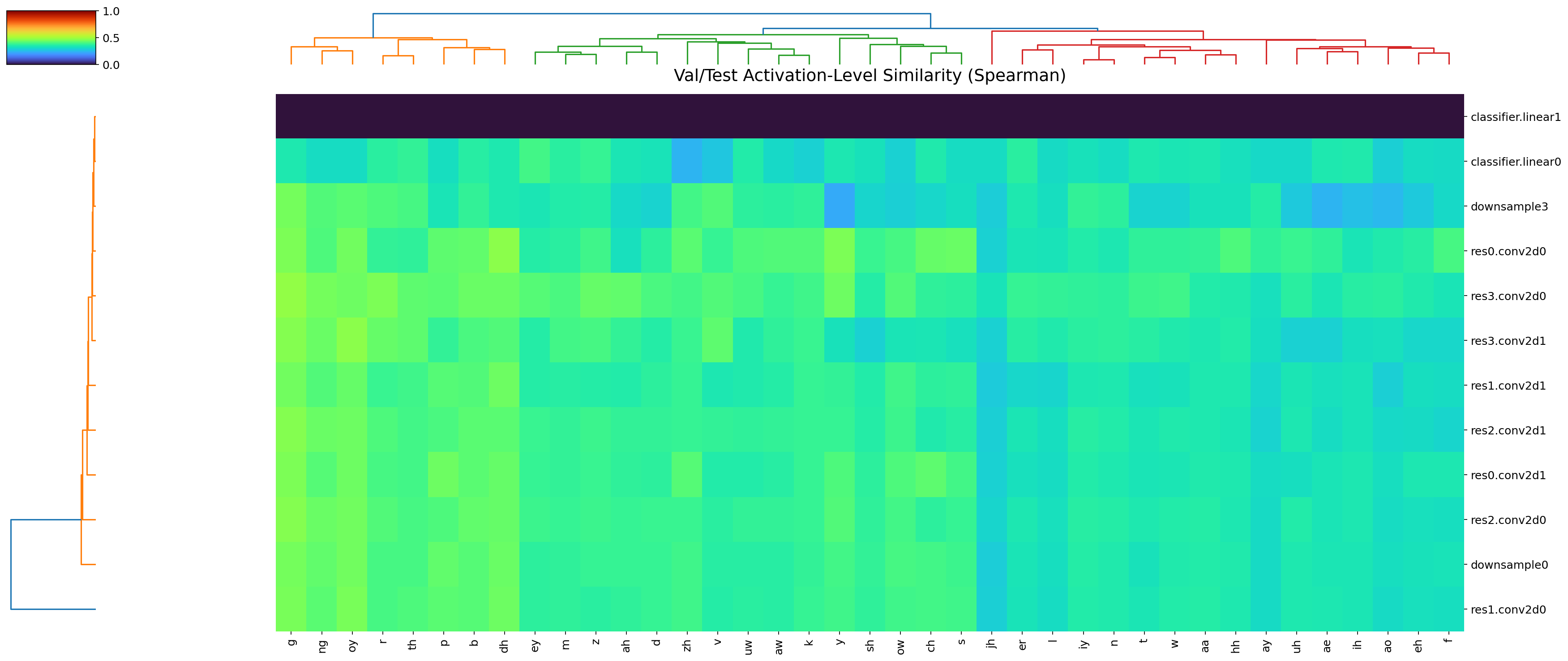}
    \caption{STFT CNN}
  \end{subfigure}\hfill
  \begin{subfigure}[t]{0.32\linewidth}
    \centering
    \includegraphics[width=\linewidth]{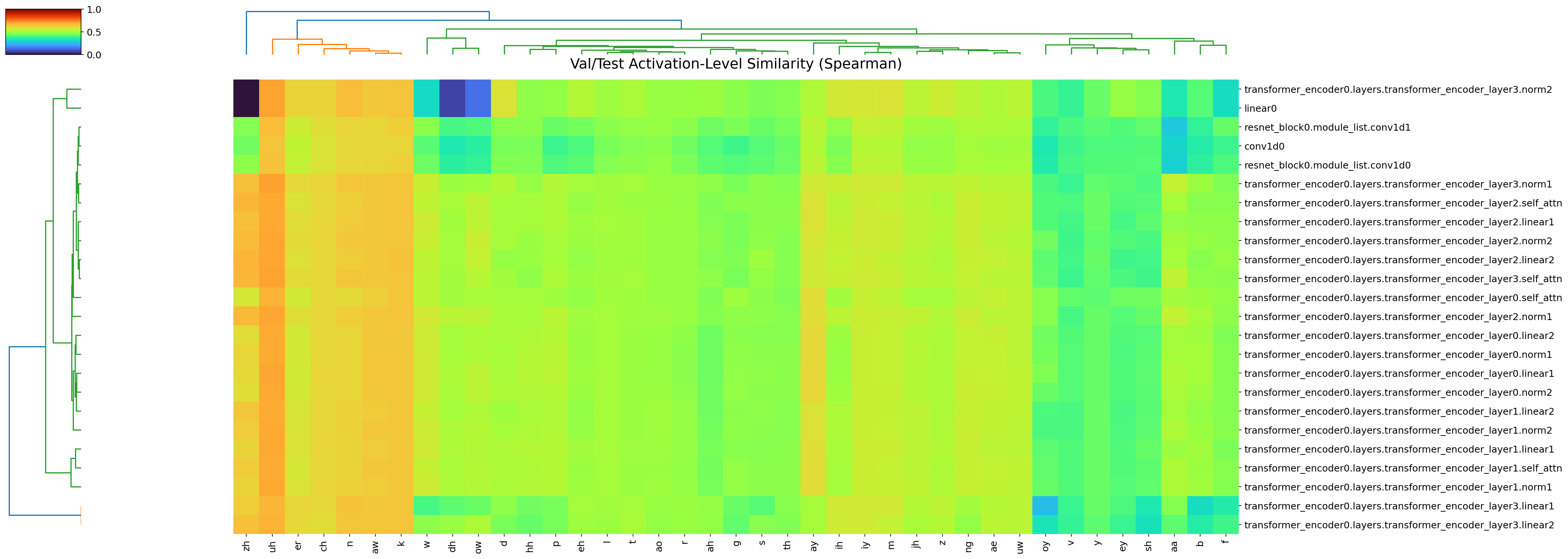}
    \caption{CNN--Transformer}
  \end{subfigure}

  \vspace{0.5em}

  \textbf{With InstanceNorm}\par\smallskip
  \textit{Pearson}\par\medskip
  \begin{subfigure}[t]{0.32\linewidth}
    \centering
    \includegraphics[width=\linewidth]{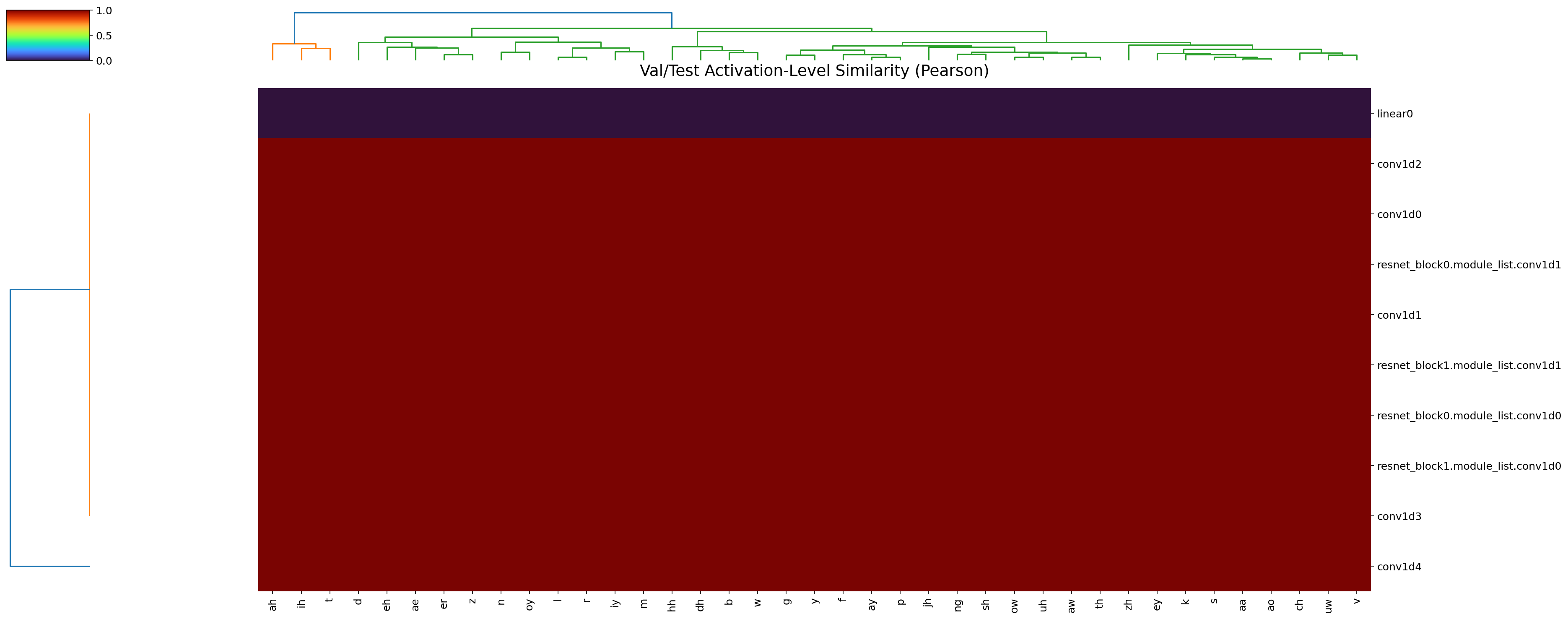}
    \caption{ResNet CNN + IN}
  \end{subfigure}\hfill
  \begin{subfigure}[t]{0.32\linewidth}
    \centering
    \includegraphics[width=\linewidth]{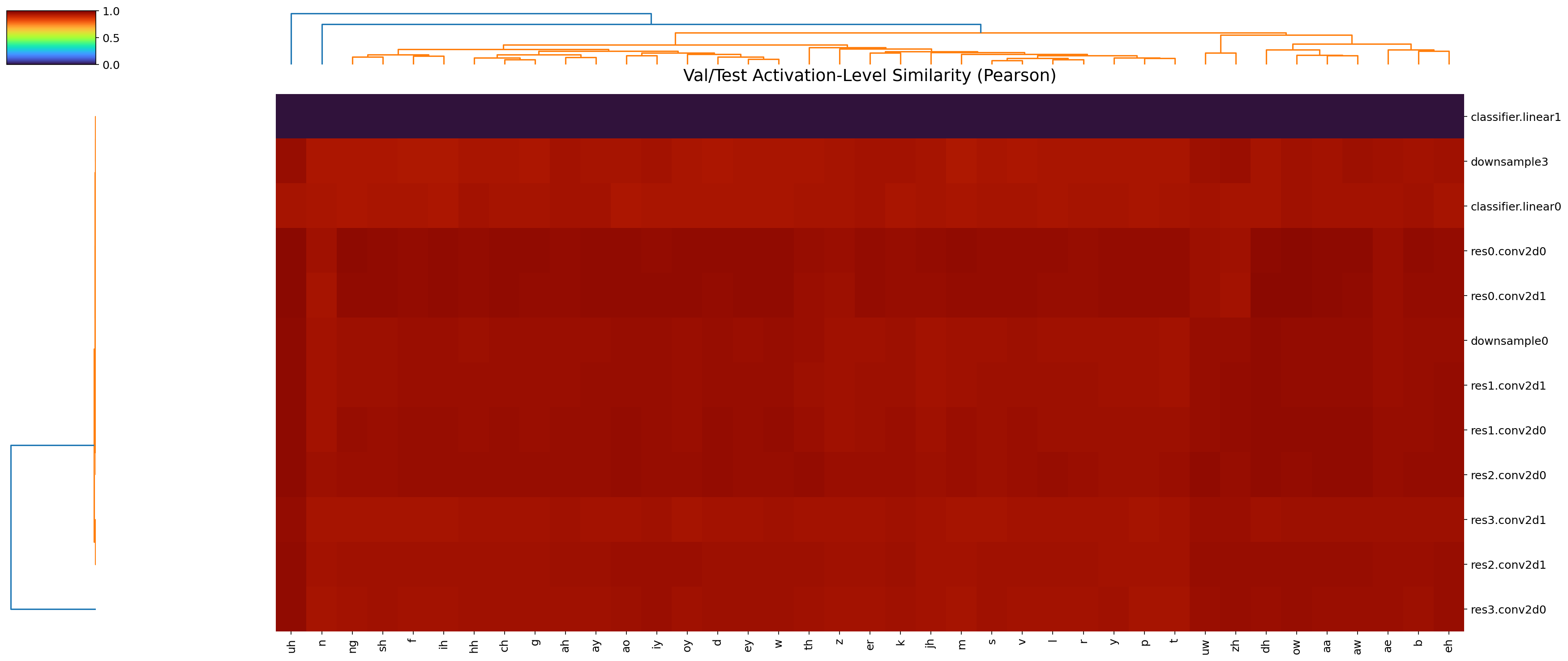}
    \caption{STFT CNN + IN}
  \end{subfigure}\hfill
  \begin{subfigure}[t]{0.32\linewidth}
    \centering
    \includegraphics[width=\linewidth]{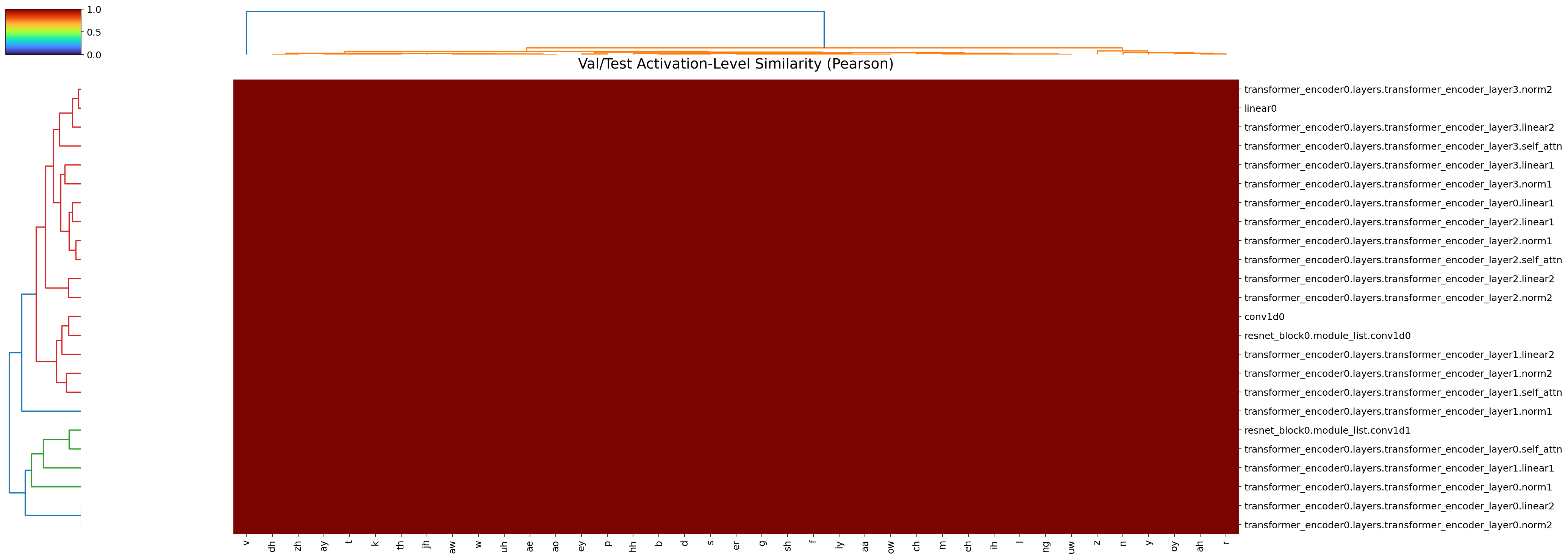}
    \caption{CNN--Transformer + IN}
  \end{subfigure}

  \medskip

  \textit{Spearman}\par\medskip
  \begin{subfigure}[t]{0.32\linewidth}
    \centering
    \includegraphics[width=\linewidth]{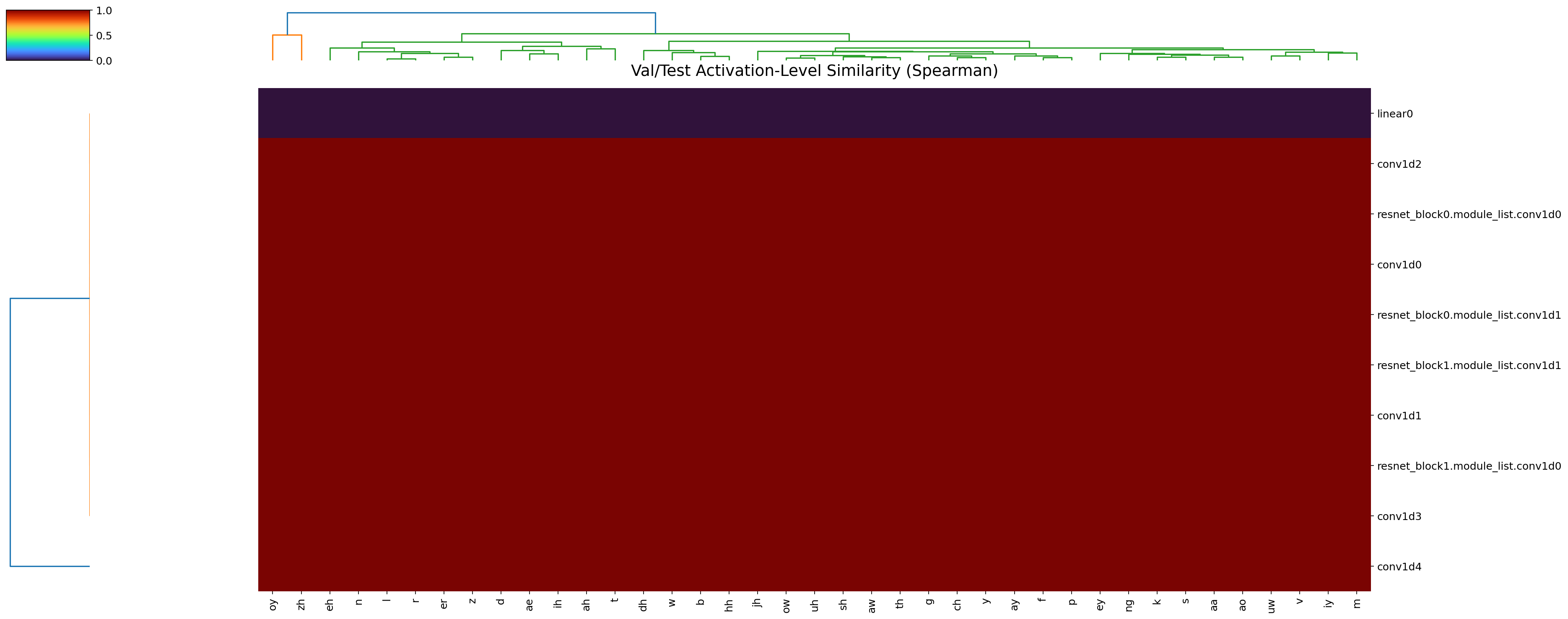}
    \caption{ResNet CNN + IN}
  \end{subfigure}\hfill
  \begin{subfigure}[t]{0.32\linewidth}
    \centering
    \includegraphics[width=\linewidth]{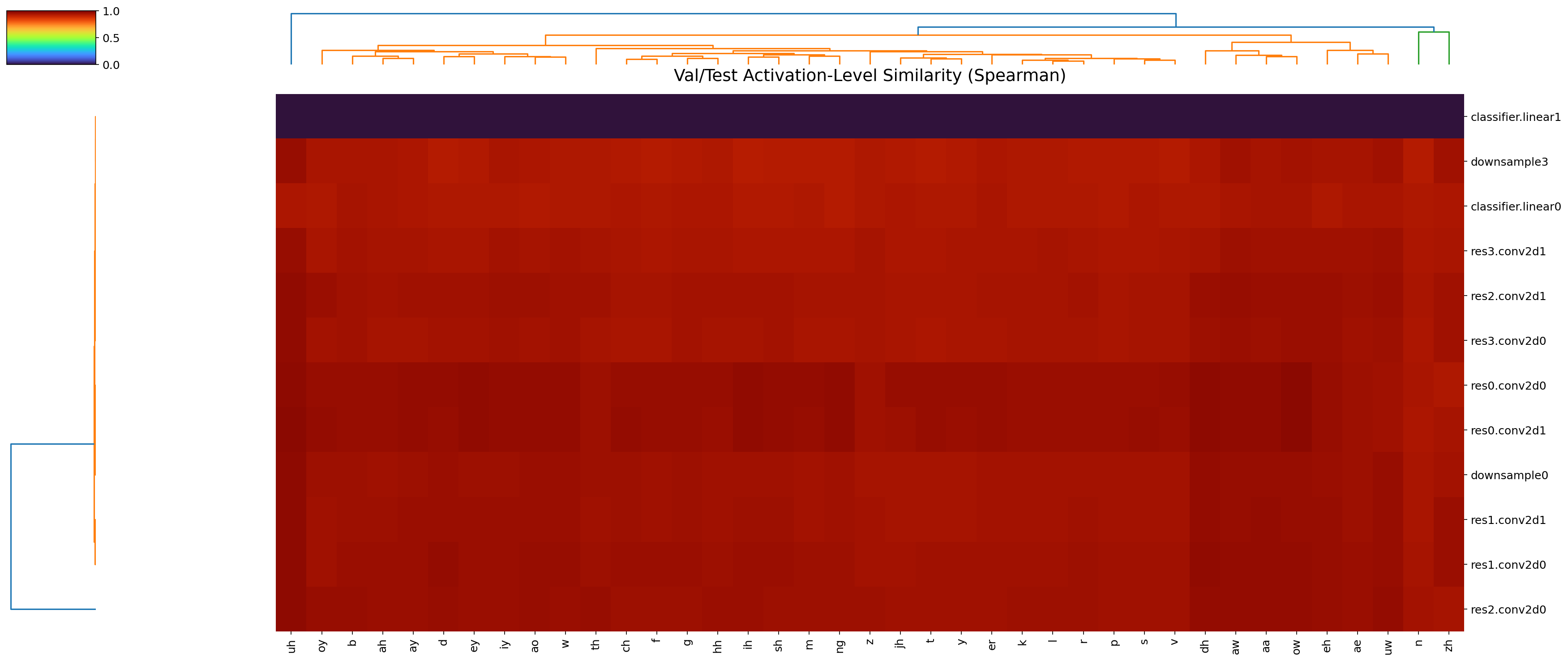}
    \caption{STFT CNN + IN}
  \end{subfigure}\hfill
  \begin{subfigure}[t]{0.32\linewidth}
    \centering
    \includegraphics[width=\linewidth]{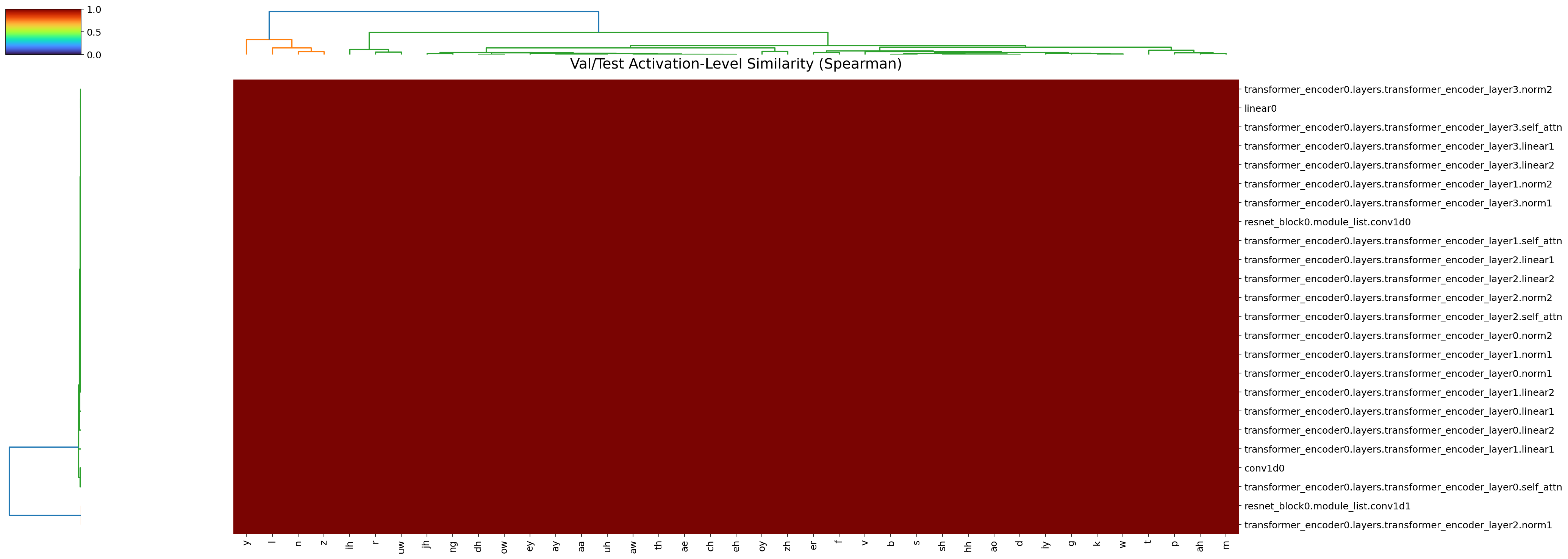}
    \caption{CNN--Transformer + IN}
  \end{subfigure}

  \caption{Layer-by-phoneme validation-test saliency similarity matrices for the three models with both standard and InstanceNorm variants. In each 3-column block, the upper row shows Pearson correlation and the lower row shows Spearman correlation. Brighter cells indicate stronger agreement between validation and test saliency patterns for the same layer-phoneme pair.}
  \label{fig:saliency_similarity_grid}
\end{figure}

To quantify the cross-split stability of saliency maps, we compute both Pearson and Spearman correlation between the validation and test saliency values for each layer-phoneme pair. A generalizing model like MEGConformer would result in highly correlated saliency map as in Figure~\ref{fig:saliency_similarity_conformer}. With MEGConformer as a strong baseline, Table~\ref{tab:saliency_similarity} and Figure~\ref{fig:saliency_similarity_grid} reveals clear differences in cross-split stability for models with or without Instance Normalization.

\begin{table}[h]
  \centering
  \caption{Validation-Test Saliency Map Similarity Summary}
  \label{tab:saliency_similarity}
  \begin{tabular}{lcccc}
    \toprule
    \textbf{Model} & \textbf{Pearson ($\mu\pm\sigma$)} & \textbf{Spearman ($\mu\pm\sigma$)} \\
    \midrule
    CNN & 0.5465 $\pm$ 0.0529 & 0.5333 $\pm$ 0.0520 \\
    CNN + InstanceNorm & 0.9996 $\pm$ 0.0003 & 0.9996 $\pm$ 0.0003 \\
    CNN--Transformer & 0.4701 $\pm$ 0.0868 & 0.5099 $\pm$ 0.0808 \\
    CNN--Transformer + InstanceNorm & 0.9999 $\pm$ 0.0000 & 0.9999 $\pm$ 0.0002 \\
    STFT CNN & 0.3770 $\pm$ 0.0448 & 0.3645 $\pm$ 0.0448 \\
    STFT CNN + InstanceNorm & 0.9562 $\pm$ 0.0102 & 0.9503 $\pm$ 0.0123 \\
    \citep{dezuazo2025megconformer} & 0.9999 $\pm$ 0.0000 & 0.9999 $\pm$ 0.0001 \\
    \bottomrule
  \end{tabular}
\end{table}

For models without Instance Normalization, the correlation of each phoneme and each layer is low (<0.6), which indicates that the saliency patterns are not stable across validation and test splits. Within these models, one can also observe the trend of higher correlation resulting in better test performance, with the ResNet CNN showing higher correlation and better test performance than the CNN--Transformer, and the STFT CNN showing the lowest correlation and worst test performance.

By contrast, the models with Instance Normalization show near-perfect correlation (>0.95) across all layers and phonemes, indicating that the saliency patterns and thus the model behavior are highly stable across splits. The relatively weaker generalizability of STFT CNN can again be seen in the slightly lower correlation scores compared to the other two InstanceNorm models, although all three are much more stable than the non-InstanceNorm variants.

The findings can be summarized two-fold: (1) the cross-split stability of saliency maps is closely associated with generalizability, and (2) Instance Normalization is a key factor in achieving that stability. The saliency map analysis therefore provides a qualitative and quantitative lens for understanding the robustness improvements observed in the benchmark results.

%% file: nstc/conclusion.tex
\section{Conclusion and Future Work}
\label{sec:conclusion}

\subsection{Core Findings}

This study systematically investigated the robustness of phoneme classification using non-invasive MEG. Through extensive ablation experiments focusing on preprocessing strategies, normalization methods, and neural architectures, we conclude that in non-invasive neural decoding tasks, data configuration and normalization strategies exert a significantly greater influence on generalization than the incremental complexity of model architectures. Specifically, Instance Normalization proved to be a critical technique for overcoming the distribution shift caused by statistical discrepancies between validation and test sets. Furthermore, while group averaging combined with repeated grouping strategies sacrifices some real-world relevance for single-trial applications, it substantially improves the signal-to-noise ratio (SNR), enabling deep learning models to extract discriminative phoneme features from highly noisy MEG signals. Stability analysis via saliency maps further confirms from an interpretability perspective that high-performing models, such as MEGConformer, maintain consistent neural sensitivity distributions across different data splits rather than merely memorizing training-specific artifacts.

\subsection{Limitation of Group Averaging}

Although group averaging significantly improved benchmark scores, this strategy relies on known categories and multiple repeated samples, creating a substantial disconnect from real-world Brain-Computer Interface (BCI) scenarios. In natural speech communication, phonemes arrive sequentially, and multiple recordings of the same stimulus are typically unavailable; therefore, an over-reliance on group averaging may obscure the fundamental deficiencies of models when processing low-SNR single-trial inputs. 

\subsection{Future Research Directions}

To address the distribution shift challenges and generalization limits identified in this study, future research should focus on the following three dimensions:

\paragraph{Enhancing robustness for single-trial decoding.}
While group averaging significantly improves benchmark scores by increasing the signal-to-noise ratio, this strategy relies on multiple repeated samples, creating a disconnect from real-world Brain-Computer Interface (BCI) scenarios. In natural speech communication, phonemes arrive sequentially, making multiple recordings of the same stimulus typically unavailable. Future work should focus on denoising techniques that can process low-SNR single-trial inputs to overcome the performance degradation observed when processing raw, non-averaged signals.

\paragraph{Developing adaptive normalization mechanisms for statistical shift.}
This study confirmed that distribution shift induced by different normalization statistics is a major obstacle to generalization. Future research should explore more adaptive normalization strategies that allow models to automatically calibrate to local data distributions during inference, ensuring consistent feature representations even when global statistics differ.

\paragraph{Learning invariant neural representations via stability constraints.}
Saliency map analysis revealed that robust models, such as MEGConformer, maintain stable and distributed phoneme-sensitive patterns across different data splits. In contrast, weaker models exhibit more concentrated or repetitive patterns that fail to generalize. Future research could incorporate stability-based objectives or invariance constraints during training to encourage the model to lock onto core neural mechanisms of language rather than memorizing training-specific statistical artifacts.

%% file: nstc/availability.tex
\section{Data Availability}
\label{sec:data_availability}

The experiments in this study use the LibriBrain phoneme-classification benchmark and the evaluation protocol of the 2025 PNPL competition \citep{ozdogan2025libribrain,landau2025pnpl}. The underlying MEG recordings and official benchmark splits are available from the original LibriBrain and PNPL resources under their respective access conditions. We did not collect any new human-subject data for this work.

Our implementation, training and evaluation scripts, and the experiment configurations used in this report are available at GitHub\footnote{\url{https://github.com/Dogeon188/libribrain-phoneme}}. Because the hidden holdout split is managed by the competition organizers, only organizer-released labels and scores can be shared for that portion of the benchmark. Derived result tables and manuscript figures are included in this report and can be reproduced from the released code, subject to access to the original LibriBrain data.

%% file: nstc/full_results.tex
\section{Extended Results}
\label{app:full_results}

Table~\ref{tab:full_results} summarizes all configurations evaluated in our ablation study. Entries with dashes indicate settings that were not submitted to the holdout evaluation due to time constraints.

For brevity, the backbone column only distinguishes between the three main architectures s stated in \ref{subsec:model_architecture}, that is, (a) the ResNet-style CNN backbone, (b) the STFT-based CNN, and (c) the CNN--Transformer hybrid. The note column indicates additional details about the configuration, such as the use of data augmentation or PanPhon targets. The final two rows list MEGConformer as external reference results rather than part of our ablation series, because those runs use a different normalization and training recipe from the models developed in this work.

Entries in the table are sorted by the holdout performance, which is the most relevant metric for our analysis. For entries with no holdout performance presented, we sort by test performance instead. The baseline configuration is underlined for easy reference. The best-performing configuration is highlighted in bold.

\begin{table}[h]
  \centering
  \setlength{\tabcolsep}{2pt}
  \scriptsize
  \caption{Extended comparison of model and data configurations. Performance is reported in F1-macro (\%).}
  \label{tab:full_results}
  \begin{tabular}{lccccccccc}
    \toprule
    \textbf{Train} & \textbf{Validation} & \textbf{Test} & \textbf{Holdout} & \textbf{Backbone} & \textbf{GroupAvg.} & \textbf{LabelBal.} & \textbf{Repeat} & \textbf{Note} \\
    \midrule
    90.81 & \textbf{71.95} & 47.47 & \textbf{35.40} & (a) & \checkmark & \checkmark & 10 &  \\
    96.62 & 44.49 & 43.17 & 24.40 & (a) & \checkmark & \checkmark & 1 & +LayerNorm \\
    80.06 & 71.77 & 44.28 & 21.60 & (a) & \checkmark & \checkmark & 1 &  \\
    48.55 & 49.31 & 34.03 & 18.80 & (a) & \checkmark & \xmark & 1 & +Augmentation\\
    \underline{34.55} & \underline{45.08} & \underline{39.53} & \underline{13.20} & (a) & \xmark & \xmark & 1 & \textbf{Baseline} \\
    68.16 & 64.27 & 47.74 & 3.60 & (a) & \checkmark & \checkmark & 1 & +Augmentation \\
    \textbf{99.43} & 62.75 & \textbf{60.95} & -- & (a) & \checkmark & \checkmark & 10 & +InstNorm \\
    83.60 & 57.72 & 55.87 & -- & (c) & \checkmark & \checkmark & 1 & +InstNorm \\
    78.42 & 70.09 & 45.02 & -- & (a) & \checkmark & \checkmark & 3 &  \\
    91.50 & 44.31 & 42.84 & -- & (a) & \checkmark & \checkmark & 1 & +BatchNorm \\
    62.63 & 43.62 & 15.91 & -- & (b) & \checkmark & \checkmark & 5 & \(N_{\text{fft}}=25, H=5\) \\
    62.05 & 40.00 & 15.78 & -- & (b) & \checkmark & \checkmark & 5 & \(N_{\text{fft}}=50, H=5\)  \\
    86.54 & 36.89 & 25.90 & -- & (b) & \checkmark & \checkmark & 5 & \(N_{\text{fft}}=25, H=5\), +InstNorm \\
    41.97 & 30.01 & 8.98 & -- & (b) & \checkmark & \checkmark & 5 & \(N_{\text{fft}}=25,H=20\)  \\
    74.37 & 64.86 & 4.29 & -- & (c) & \checkmark & \checkmark & 1 &  \\
    -- & 51.68 & 3.33 & -- & (c) & \checkmark & \checkmark & 1 & Ternary PanPhon  \\
    75.22 & 63.82 & 0.98 & -- & (c) & \checkmark & \checkmark & 1 &  \\
    -- & 57.21 & 0.88 & -- & (c) & \checkmark & \checkmark & 1 & Binary PanPhon  \\
    -- & -- & 0.06 & -- &  (c) & \checkmark & \checkmark & 1 & w/ Dist. Mapper  \\
    -- & -- & 0.12 & -- & (c)& \checkmark & \checkmark & 1 & w/ Dist. Mapper  \\
    \midrule
    61.10 & 51.55 & 49.99 & -- & \citep{dezuazo2025megconformer} & -- & -- & -- & 10 epochs \\
    68.55 & 65.67 & \textbf{64.09} & -- & \citep{dezuazo2025megconformer} & -- & -- & -- & 36 epochs \\
    \bottomrule
  \end{tabular}
\end{table}
